\documentclass[twocolumn]{aastex61}
\usepackage{graphics,graphicx,natbib,subfigure,wrapfig}
\usepackage{ textcomp }
\usepackage{amssymb}

\newcommand{\ee}[1]{\mbox{${} \times 10^{#1}$}}
\newcommand{\eten}[1]{\mbox{$10^{#1}$}}
\newcommand{\msun}{\mbox{M$_\odot$}}

\newcommand{\water}{H$_2$O}

\newcommand{\cotwo}{CO$_{2}$}

\newcommand{\hh}{\mbox{{\rm H}$_2$}}
\newcommand{\hhhp}{\mbox{{\rm H}$_3^+$}}
\newcommand{\nn}{\mbox{{\rm N}$_2$}}
\newcommand{\ammonia}{\mbox{{\rm NH}$_3$}}

\newcommand{\hep}{He$^+$}
\newcommand{\methanol}{\mbox{{\rm CH$_3$OH}}}

\begin{document}

\title{Unlocking CO Depletion in Protoplanetary Disks I. The Warm Molecular Layer}

\correspondingauthor{Kamber R. Schwarz}
\email{kamberrs@umich.edu}

\author{Kamber R. Schwarz}
\affiliation{Department of Astronomy, University of Michigan, 1085 South University Ave., Ann Arbor, MI 48109, USA}
\author{ Edwin A. Bergin}
\affiliation{Department of Astronomy, University of Michigan, 1085 South University Ave., Ann Arbor, MI 48109, USA}
\author{L. Ilsedore Cleeves}
\affiliation{Harvard-Smithsonian Center for Astrophysics, 60 Garden Street, Cambridge, MA 02138, SA},
\author{Ke Zhang}
\affiliation{Department of Astronomy, University of Michigan, 1085 South University Ave., Ann Arbor, MI 48109, USA}
\author{Karin I. \"{O}berg}
\affiliation{Harvard-Smithsonian Center for Astrophysics, 60 Garden Street, Cambridge, MA 02138, SA},
\author{Geoffrey A. Blake}
\affiliation{Division of Geological \& Planetary Sciences, MC 150-21, California Institute of Technology, 1200 E California Blvd, Pasadena, CA 91125}
\author{Dana Anderson}
\affiliation{Division of Geological \& Planetary Sciences, MC 150-21, California Institute of Technology, 1200 E California Blvd, Pasadena, CA 91125}

\begin{abstract}
CO is commonly used as a tracer of the total gas mass in both the interstellar medium and in protoplanetary disks. Recently there has been much debate about the utility of CO as a mass tracer in disks.
Observations of CO in protoplanetary disks reveal a range of CO abundances, with measurements of low CO to dust mass ratios in numerous systems. 
One possibility is that carbon is removed from CO via chemistry.
However, the full range of physical conditions conducive to this chemical reprocessing is not well understood. 
We perform a systematic survey of the time dependent chemistry in protoplanetary disks for 198 models with a range of physical conditions. We varying dust grain size distribution, temperature, comic ray and X-ray ionization rate, disk mass, and initial water abundance, detailing what physical conditions are necessary to activate the various CO depletion mechanisms in the warm molecular layer. We focus our analysis on the warm molecular layer in two regions: the outer disk (100 au) well outside the CO snowline and the inner disk (19 au) just inside the midplane CO snow line.
After 1 Myr, we find that the majority of models have a CO abundance relative to \hh\ less than \eten{-4} in the outer disk, while an abundance less than \eten{-5} requires the presence of cosmic rays.
Inside the CO snow line, significant depletion of CO only occurs in models with a high cosmic ray rate. If cosmic rays are not present in young disks it is difficult to chemically remove carbon from CO. Additionally, removing water prior to CO depletion impedes the chemical processing of CO. 
Chemical processing alone cannot explain current observations of low CO abundances. Other mechanisms must also be involved.
\end{abstract}


\section{Introduction}
CO is arguably the mostly commonly observed molecule in astronomy, being both highly abundant and highly emissive. In molecular clouds it is used as a tracer of the total gas mass, with an observed abundance relative to \hh\ of $\sim 0.5 - 3\ee{-4}$ \citep{Bergin17}. This abundance ratio is often assumed to hold for protoplanetary disks. However, as early as the first detection of CO in a protoplanetary disk there has been a discrepancy between the gas mass inferred from CO and that estimated from the dust \citep{Dutrey94}. Recent surveys of protoplanetary disks with ALMA have revealed that many systems 
have a low CO-to-dust mass ratio compared to the ISM \citep[e.g.,][]{Ansdell16,Long17}. 
This has been attributed to either a low total gas mass or a low abundance of CO relative to \hh\ \citep{Williams14,Miotello17}. 
Distinguishing between the two scenarios requires an alternative tracer of the total gas mass, such as HD, which has been detected in three systems so far \citep{Bergin13,McClure16}. For all three systems CO is under-abundant, with abundances relative to \hh\ in the range $\eten{-6}-2\ee{-5}$  \citep{Favre13,Schwarz16,McClure16}. 
High spatial resolution observations of CO isotopes in TW Hya reveal that CO is also under-abundant interior to the midplane CO snow line \citep{Zhang17}. In this source simple freeze out of CO onto grains is insufficient to explain the observations.
Further, CO is not the only under-abundant volatile. TW Hya is under-abundant in gas phase atomic carbon \citep{Kama16} and \water\ vapor is depleted in numerous systems, including TW Hya \citep{Du15,Du17}. 

Previous modeling has explored the affects of photodissociation, freeze-out onto dust grains, and isotope selective self-shielding on the observed CO abundances.
\citet{Williams14} apply models which correct for a reduction in gas phase CO due to both photodissociation and freeze out to observations of disks in Taurus. Even when accounting for these effects they find CO to dust ratios well below the those measured in the ISM.
CO abundance measurements are often based on observations of optically thin emission from less abundant isotopologues, using isotopologue abundance ratios to convert to a total CO abundance.
\citet{Miotello14} demonstrate that changes in isotopologue abundance ratios due to self-shielding can lead to underestimates of the CO abundance by up to an order of magnitude, but cannot explain the CO abundances in the most depleted systems. Further, the CO abundances relative to \hh\ based on detections of HD exclude any gas below 20 K. This is because CO freezes onto dust grains at temperatures below 20 K while the emissivity of the HD 1-0 drops sharply for temperatures less than 20 K \citep{Favre13}. Taken together, these studies show that even in the warm molecular layer CO is under-abundant.

Several mechanisms have been put forth to explain the observed low CO abundances, including gas disk dispersal \citep{Bai16}, dust dynamics, grain growth, settling, dust drift \citep{Salyk08,Krijt16,Xu17}, and chemical reprocessing.
Multiple studies have shown that CO can be depleted via chemical pathways within the disk \citep{Bergin14,Reboussin15,Eistrup16,Yu16}. However, to date these studies have been limited in their scope, lacking a holistic view of the disk conditions for which different chemical pathways are active as well as how the efficiency of such pathways vary from disk to disk.
In particular, it is unclear if such mechanisms are efficient enough to fully explain the observed CO abundances without invoking other mechanisms such as grain growth.

Most chemical models require a source of ionization, with the carbon eventually converted to \cotwo\ ice, though other pathways have been identified. 
In the chemical models of \citet{Bergin14}, with physical parameters based on TW Hya, CO gas reacting with \hep\ results in much of the carbon being reprocessed into hydrocarbons, a mechanism first put forth by \citet{Aikawa99}. These hydrocarbons then freeze onto grains.
\citet{Reboussin15} explored a grid of chemical models spanning a range of disk temperatures and initial chemical abundances. In their models CO is depleted by an order of magnitude for temperatures less than 30 K, going primarily into \cotwo\ ice. In these models the reprocessing of CO ice is the result of cosmic-ray induced UV photons. As such, the final ice composition depends on the incident cosmic ray rate assumed.
\citet{Eistrup16} focus on the midplane chemistry inside of 30 au for a single disk physical model. They find that much of the CO in this region is converted to \cotwo\ ice, though at 30 au a significant fraction of the carbon is found in CH$\mathrm{_4}$ ice as well.
\citet{Yu16} explore the chemical effects in a warmer disk, heated by both stellar radiation and viscous heating, with an enhanced gas-to-dust ratio of 1000 compared to the commonly assumed value of 100. In their model disk the midplane temperature is greater than the CO freeze out temperature interior to 70 au. However, CO is still depleted in the presence of cosmic rays, with the carbon found in \cotwo\ ice as well as ices composed of complex organics.

Clearly it is possible to chemically deplete CO in protoplanetary disks under a variety of physical conditions. 
The depletion mechanisms identified in previous work require the presence of ionizing radiation, primarily cosmic rays, and/or warm disk temperatures, which keep CO in the gas phase. 
 However, none of the previous studies systematically explore how different dust populations impact the chemistry.
Furthermore, there is doubt as to whether high ionization rates are present in protoplanetary disks.  In our own solar system the solar wind modulates cosmic rays within the heliosphere \citep{Gleeson68}. Similarly, \citet{Cleeves13} argue that winds from T-Tauri stars can modulate the incident cosmic ray rate for the surrounding disk. Indeed, the N$_{2}$H$^+$ emission observed in TW Hya is consistent with a low cosmic ray ionization rate \citep{Cleeves14}. 
Given the expected ubiquity of stellar winds, sub-ISM cosmic ray levels should be the norm among protoplanetary disks, removing the most commonly invoked mechanism for CO depletion. 

In this paper we aim to systematically constrain the subset of physical conditions necessary to enable chemical reprocessing of CO. To this end we explore the chemical evolution for a large grid of physical models, varying the cosmic ray ionization rate, the incident X-ray flux, the mass of the disk, the temperature of the disk, the fraction of dust mass in large grains. All of these parameters affect either the ionization or temperature structure. Additionally, we vary the the amount of water initially available to the chemical network in order to explore how oxygen depletion impacts the carbon chemistry. 
Current observations of low volatile abundances primarily probe the warm molecular layer, where volatile elements such as carbon are primarily in gas phase molecules \citep{Aikawa99}. In this paper we focus on the conditions and timescales needed to chemically remove CO from the warm molecular layer. Future work will explore the results for the midplane and the implications for planet formation.
\S \ref{model} details our modeling framework and describes the parameter space covered by our grid of models. The results of our model grid are explored in \S \ref{results}. In \S \ref{discussion} we compare our results to previous work and discuss the implications for observations of volatile carbon in protoplanetary disks. Finally, we summarize our findings in \S \ref{summary}.

\section{Model}\label{model}
\begin{figure}[ht!]
\setlength{\intextsep}{0pt}
    \includegraphics[width=\columnwidth]{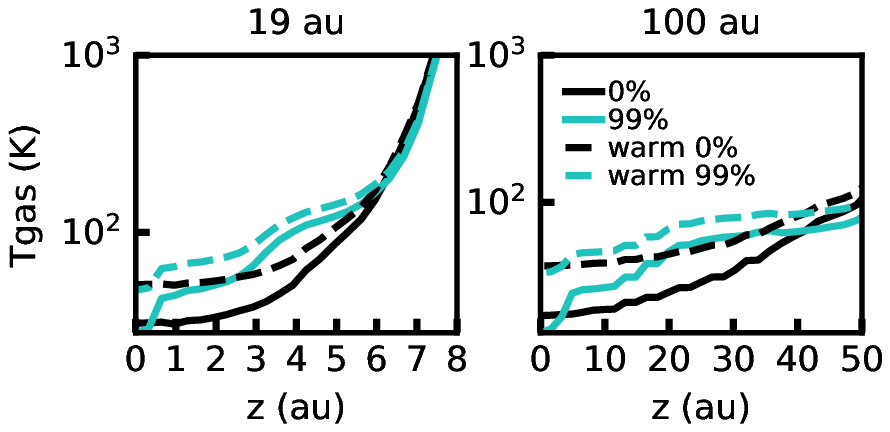}
\caption{Vertical gas temperature profiles at 19 au and 100 au for the 0.03 \msun\ disk. The 0\% large grain model is shown in black and the 99\% large grain model is shown in grey. Dashed lines indicate the temperature in the warm disk models. \label{mtemp}}
\end{figure}

The physical model is a two-dimensional, azimuthally symmetric disk generated using the radiative transfer code TORUS \citep{Harries00}, which calculates the temperature structure for a given distribution of gas and dust (Figure~\ref{mtemp}). We consider disk masses of 0.1, 0.03, and 0.003 \msun\ with an inner radius of 0.1 au and an outer radius of 200 au.
The surface density distribution for the gas and the small grains ($r_d = 0.005 - 1$ \micron) is:
\begin{equation}
\Sigma_{g}(R)=\Sigma_{c}\left(\frac{R}{R_{c}}\right)^{-\gamma} \exp{\left[-\left(\frac{R}{R_{c}}\right)^{2-\gamma}\right]},
\end{equation}
where $\Sigma_{c}$ is the disk surface density at $R_{c} = 100$ au and $\gamma=1$. 
The density distribution for the small grains is
\begin{equation}
\rho_{s} = \frac{(1-f)\Sigma}{\sqrt{2\pi}R h} \exp{\left[-\frac{1}{2}\left(\frac{Z}{H}\right)^{2}\right]}.
\end{equation}
For the large grains ($r_d = 0.005 - 1000$ \micron):
\begin{equation}
\rho_{l} = \frac{f\Sigma}{\sqrt{2\pi}R\chi h} \exp{\left[-\frac{1}{2}\left(\frac{Z}{\chi H}\right)^{2}\right]},
\end{equation}
where $f$ is the fraction of the dust mass in large grains, $Z$ is the height in the disk, and $\chi$ is the fractional scale height of the large grains. The scale height for the gas and small grains is:
\begin{equation}
h(r)=12\ \mathrm{au} \left(\frac{R}{R_{c}}\right)^{0.3}
\end{equation}
and
\begin{equation}
H = hR
\end{equation}
Both the small and large grains follow an MRN size distribution \citep{Mathis77} where $n_g \propto r_{d}^{-3.5}$ and have optical properties consistent with a population that is comprised entirely of astronomical silicates \citep{Draine84}. The overall gas-to-dust mass ratio at each radius is 100. 
The large grains are more settled than the small grains, simulating the observed segregation between large and small grains.
Balancing diffusion and settling, the fractional scale height of the large grains is:
\begin{equation}
\chi = \sqrt{\frac{\alpha}{\alpha + \Omega t_{stop}}\frac{1+\Omega t_{stop}}{1+2\Omega t_{stop}}},
\end{equation}
\citep{Youdin07} where $\alpha=\eten{-3}$ and $\Omega$ is the orbital frequency. Following the approach of \citet{Krijt16} we calculate the stopping time, the time required for the velocity of a grain to be reduced by a factor of $e$, to be 
\begin{equation}
t_{stop} = \sqrt{\frac{\pi}{8}} \frac{\rho_{g} r_{d}}{\rho_l c},
\end{equation}
where $r_{d} = 1$ mm is the dust grain radius, $\rho_{g}$ is the internal density of a grain, $\rho_{l}$ is the density of large grains, and $c$ is the speed of sound.

\begin{figure}[ht!]
\setlength{\intextsep}{0pt}
    \includegraphics[width=\columnwidth]{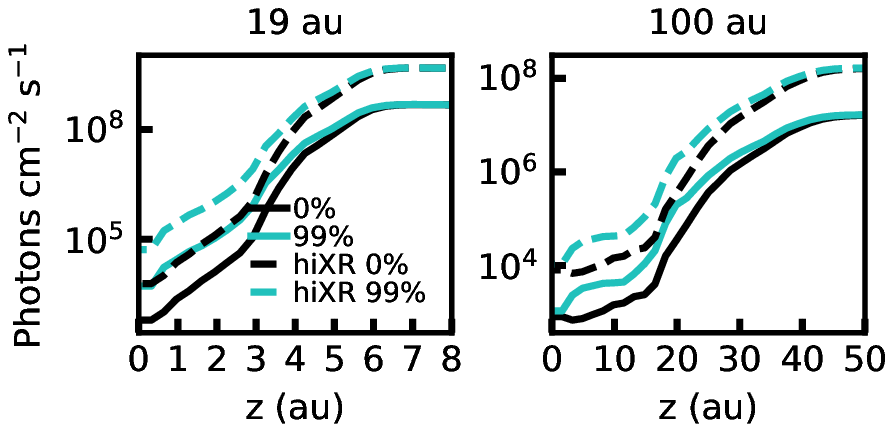}
\caption{Vertical profiles of the integrated X-ray flux at 19 au and 100 au for the 0.03 \msun\ disk. The 0\% large grain model is shown in black and the 99\% large grain model is shown in grey. Dashed lines indicate the temperature in the warm disk models. \label{mxr}}
\end{figure}

\begin{figure}[ht!]
\setlength{\intextsep}{0pt}
    \includegraphics[width=\columnwidth]{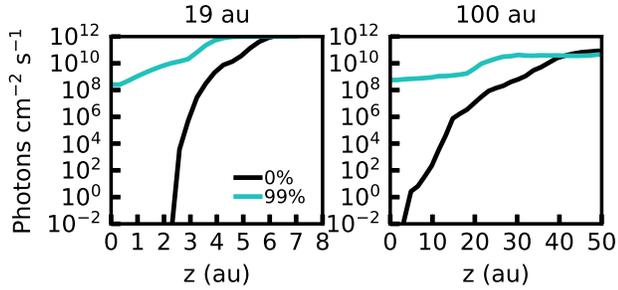}
\caption{Vertical profiles of the integrated UV flux at 19 au and 100 au for the 0.03 \msun\ disk. The 0\% large grain model is shown in black and the 99\% large grain model is shown in grey. Dashed lines indicate the temperature in the warm disk models. \label{muv}}
\end{figure}

UV and X-ray radiative transport are computed using Monte Carlo methods as described by \citet{Bethell11b,Bethell11a}. 
We consider an X-ray spectrum in the range 1-10 keV.
The overall integrated X-ray luminosity is either \eten{30} erg s$^{-1}$ or \eten{31} erg s$^{-1}$, see Figure~\ref{mxr} \citep{Cleeves15}. In comparison the X-ray luminosity of TW Hya is 2\ee{30} erg s$^{-1}$ \citep{Kastner99,Raassen09,Brickhouse10}.
We consider UV flux from the central star in the range $930 -2000$ \AA, and include scattering of Ly$\alpha$ photons by hydrogen atoms as well as dust grains, see Figure~\ref{muv} \citep{Bethell11a}.  The central star is a T Tauri star based on the properties of TW Hya.
The cosmic ray ionization rates, $\zeta_{CR}$, and the method for calculating the attenuation are taken from \citep{Cleeves13}. We consider  $\zeta_{CR}$  appropriate for the diffuse ISM \citep[2\ee{-17} s$^{-1}$,][]{Webber98} and a reduced rate consistent with the current flux in the solar system at 1 AU during solar maximum (1.6\ee{-19} s$^{-1}$), the later of which results in the best agreement with the observed molecular emission from TW Hya \citep[][]{Cleeves14}.

Table 1 lists the set of physical conditions covered by our models. For each disk mass we generate a model for a range of large grain fractions, $f$, from 0 to 0.99. Our fiducial model uses an X-ray luminosity of $\eten{30}$ erg s$^{-1}$ and a cosmic ray ionization rate of $1.6\ee{-19}$ s$^{-1}$. The high X-ray luminosity models use an X-ray luminosity of $\eten{31}$ erg s$^{-1}$ with all other parameters the same as for the fiducial models, while the high cosmic ray rate models assume $\zeta_{CR} = 2\ee{-17}$ s$^{-1}$. Additionally we consider warm disk models, in which the gas and dust temperatures have been increased by 20 K everywhere in the disk, for the fiducial and high cosmic ray models. 

\begin{deluxetable}{ll}
\tabletypesize{\scriptsize}
\tablewidth{0pt}
\tablecaption{Physical Model Properties}\label{grid}
\tablehead{
\colhead{Parameter} & \colhead{Values}  \\
}
\startdata
M$_{disk}$ ($\msun$) & 0.1, 0.03, 0.003 \\ 
L$_{XR}$ (erg s$^{-1}$) & 1E30, 1E31 \\ 
$\zeta_{CR}$ (s$^{-1}$) & 1.6E-19, 2E-17 \\
$f_l$ & 0.0, 0.1, 0.2, 0.3, 0.4, 0.5, 0.6, 0.7, 0.8, 0.9, 0.99\\
R$_{in}$ (au) & 0.1 \\
R$_{out}$ (au) & 200 \\
\enddata
\end{deluxetable}

Our chemical model is that described by \citet{Cleeves14}, based on the chemical networks of \citet{Fogel11} and \citet{Smith04} and updated to include the reaction rates from \citet{McElroy13}. The reaction network includes photodesorption, photodissociation, freeze-out onto grains, cosmic ray and X-ray ionization, gas phase ion and electron reactions, and self-shielding of \hh\ and CO. The model also includes a limited network of grain surface reactions.

\begin{deluxetable}{llllll}
\tabletypesize{\scriptsize}
\tablewidth{0pt}
\tablecaption{Initial abundances relative to \hh}\label{abundances}
\tablehead{
\colhead{Species} & \colhead{Abundance} & \colhead{Species} & \colhead{Abundance} & \colhead{Species} & \colhead{Abundance}  \\
}
\startdata
\hh\         & 1.00E00 & \ammonia  &   1.60E-07 &  SO & 1.00E-08 \\
He          &2.80E-01 & CN    &      1.20E-07  & CS    &      8.00E-09 \\
CO         & 2.00E-04 &  HCN   &      4.00E-08 &  C$^+$    &      2.00E-09 \\
\water(gr)  &  5.00E-04 & H$_3^+$   &      2.00E-08 & Si$^+$   &      2.00E-11 \\
N         &  4.50E-05 & HCO$^+$  &      1.80E-08 & S$^+$    &      2.00E-11 \\  
\nn\    &      2.00E-06 & C$_2$H   &      1.60E-08 & Mg$^+$   &      2.00E-11 \\
C     &      1.40E-06  & H$_2$CO & 1.60E-08 & Fe$^+$   &      2.00E-11 \\
\enddata
\end{deluxetable}



When calculating the chemistry, each radius is broken into 45 vertical zones. The chemistry in each zone is calculated independently, with the exception of self-shielding which considers the abundances above the current zone.
The dust size distribution in each zone factors into calculating the rates for reactions such as freeze-out and photodesorption, where the grain surface area is important.
Each zone starts with the same initial abundances relative to \hh, based on the abundances for the model molecular cloud of \citet{Aikawa99} (Table \ref{abundances}). 
Additionally, for the 0.03 \msun\ fiducial, high X-ray, and high cosmic ray models, we calculate the chemistry for a reduced oxygen abundance by setting the initial \water\ ice abundance to zero. Including these models, the total number of unique models is 198.
For each model the chemistry runs for 6 Myr.

\section{Results}\label{results}
We focus our analysis at two radii: 19 au, which is interior to the midplane CO snowline; here defined as the location in the midplane where the CO gas and CO ice abundances are equal, and 100 au, which is typical of the outer disk.
In this paper we further focus on the vertical region known as the warm molecular layer \citep{Aikawa99}, the disk region typically probed by observations. For the purpose of our analysis we define the warm molecular layer as the vertical region bounded by the vertical CO snowline and the height at which the CO gas phase abundance relative to \hh\ drops by a factor of 1/e from its peak abundance at that radius, see Figure~\ref{wmldef} for an example. 
The abundances reported in this paper are calculated by vertically integrating the absolute abundance of a given species in the warm molecular layer, then dividing the resulting column density by the \hh\ column density over the same region.

\begin{deluxetable}{rcccc}
\tabletypesize{\scriptsize}
\tablewidth{0pt}
\tablecaption{General CO Abundance Trends}\label{gentab}
\tablehead{
\colhead{} & \multicolumn2c{19 au} & \multicolumn2c{100 au} \\
\colhead{Model Type} & \colhead{Depletion} & \colhead{No Depletion} & \colhead{Depletion} & \colhead{No Depletion} 
}
\startdata
high CR & \checkmark & &  \checkmark & \\
warm high CR & & \checkmark  &  & \checkmark  \\
high X-rays & & \checkmark & \checkmark & \\
fiducial & & \checkmark & \checkmark  & \\
warm fiducial & & \checkmark & & \checkmark  \\
99\% large grains & \checkmark & & & \checkmark \\
\enddata
\end{deluxetable}

\begin{figure*}[ht!]
\setlength{\intextsep}{0pt}
    \includegraphics[]{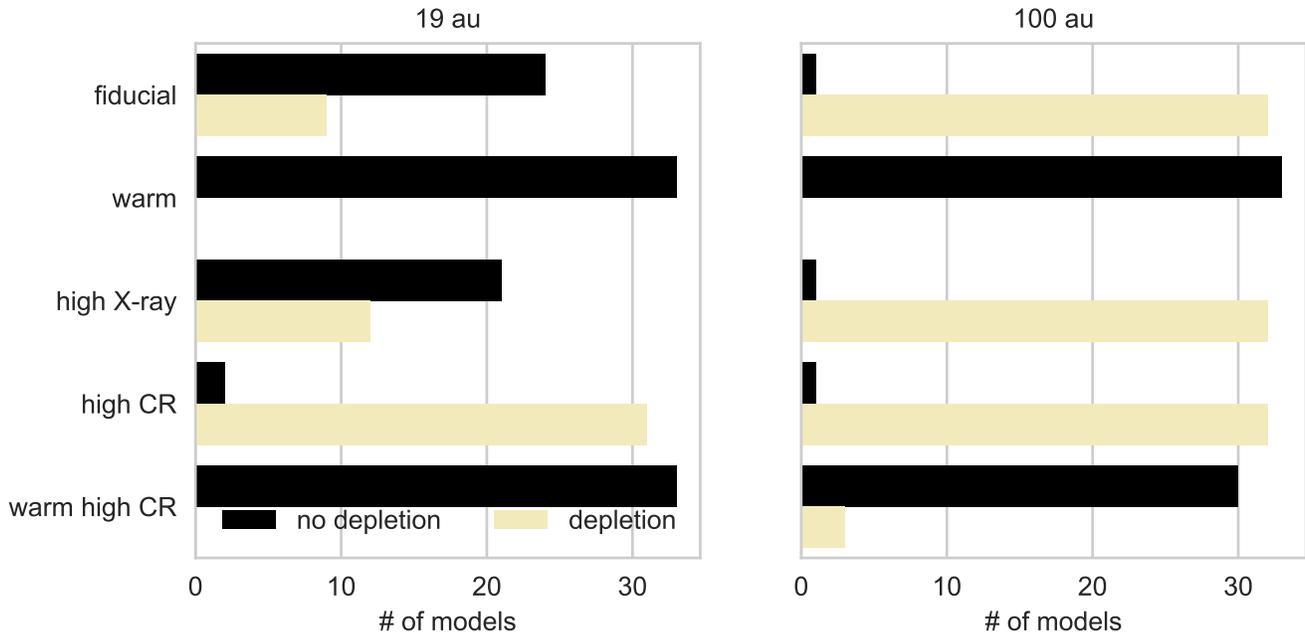}
\caption{Breakdown of the number of models that are depleted (N(CO) $<$ \eten{-4}) and not depleted (N(CO) $>$ \eten{-4}) at 1 Myr. \label{bar}}
\end{figure*}
We run a total of 165 models with our standard, ISM initial abundance. 
Figure~\ref{bar} and Table~\ref{gentab} summaries which models are able to chemically reprocess CO. 
Of these, in the warm molecular layer at 100 au, by 1 Myr 58.18\% of models show some depletion, with CO abundances between \eten{-4} and \eten{-5} while 1.82\% of models are extremely depleted, with abundances less than \eten{-5}. By 6 Myr 57.58\% are depleted and an additional 17.56\% of models are extremely depleted. 
In the inner disk at 1 Myr 18.79\% of models are depleted and 12.73\% are extremely depleted. By 6 Myr 26.06\% of models are depleted and 27.27\% are extremely depleted.
In the outer disk most models are able to reduce the CO column density by an order of magnitude, while inside the midplane snow line a high cosmic ray rate is generally required to remove carbon from CO. Below we discuss the conditions necessary for CO depletion in greater detail. 

\subsection{Outer disk}
\begin{figure}[ht!]
\setlength{\intextsep}{0pt}
    \includegraphics[width=\columnwidth]{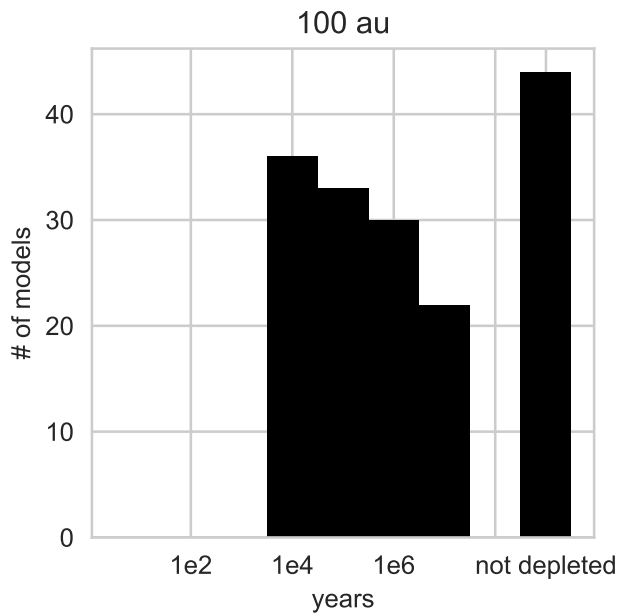}
\caption{Time at which the CO abundance first drops below \eten{-4} at 100 au for each model.\label{outtimebar}}
\end{figure}

\begin{figure}[ht!]
\setlength{\intextsep}{0pt}
    \includegraphics[width=\columnwidth]{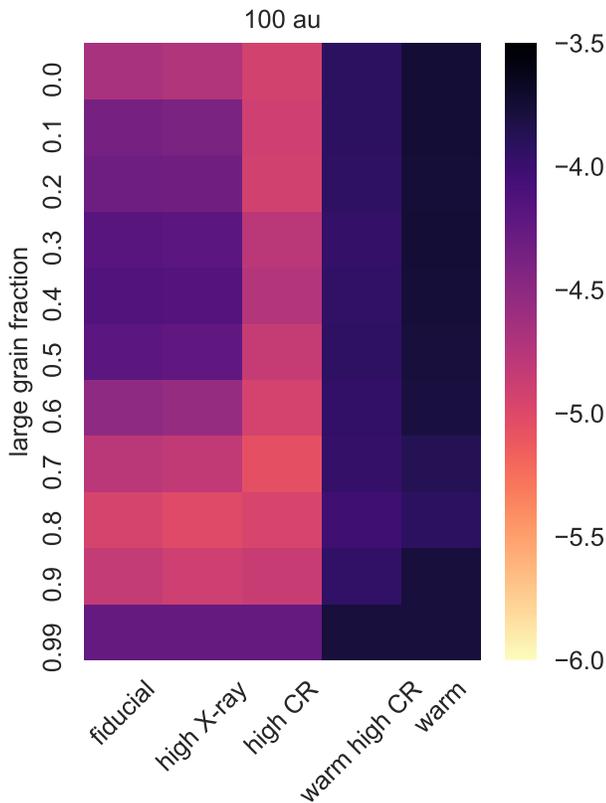}
\caption{Log CO abundance relative to \hh\ at 100 au for each model with a disk mass of 0.03 \msun. Additional figures for the 0.1 and 0.003 \msun\ disks can be found in Appendix \ref{appfigs}. \label{heatout}}
\end{figure}

In the warm molecular layer at 100 au 41.8\% of our 165 models have a CO abundance of less than \eten{-4} by 0.1 Myr (Figure~\ref{outtimebar}), including the majority of the fiducial models, high cosmic ray rate models (hiCR), and high X-ray luminosity models (hiXR). Additionally, the models with a high cosmic ray rate where the temperature has been increased by 20 K (warm hiCR) are able to deplete CO on timescales of a few million years such that 60.0\% of models are depleted by 1 Myr, increasing to 72.1\% by 3 Myr. The CO abundance in each model at 1 Myr is given in Figure~\ref{heatout} and Figure~\ref{largeheatout}.

\begin{figure*}[]
\setlength{\intextsep}{0pt}
    \includegraphics[width=\textwidth]{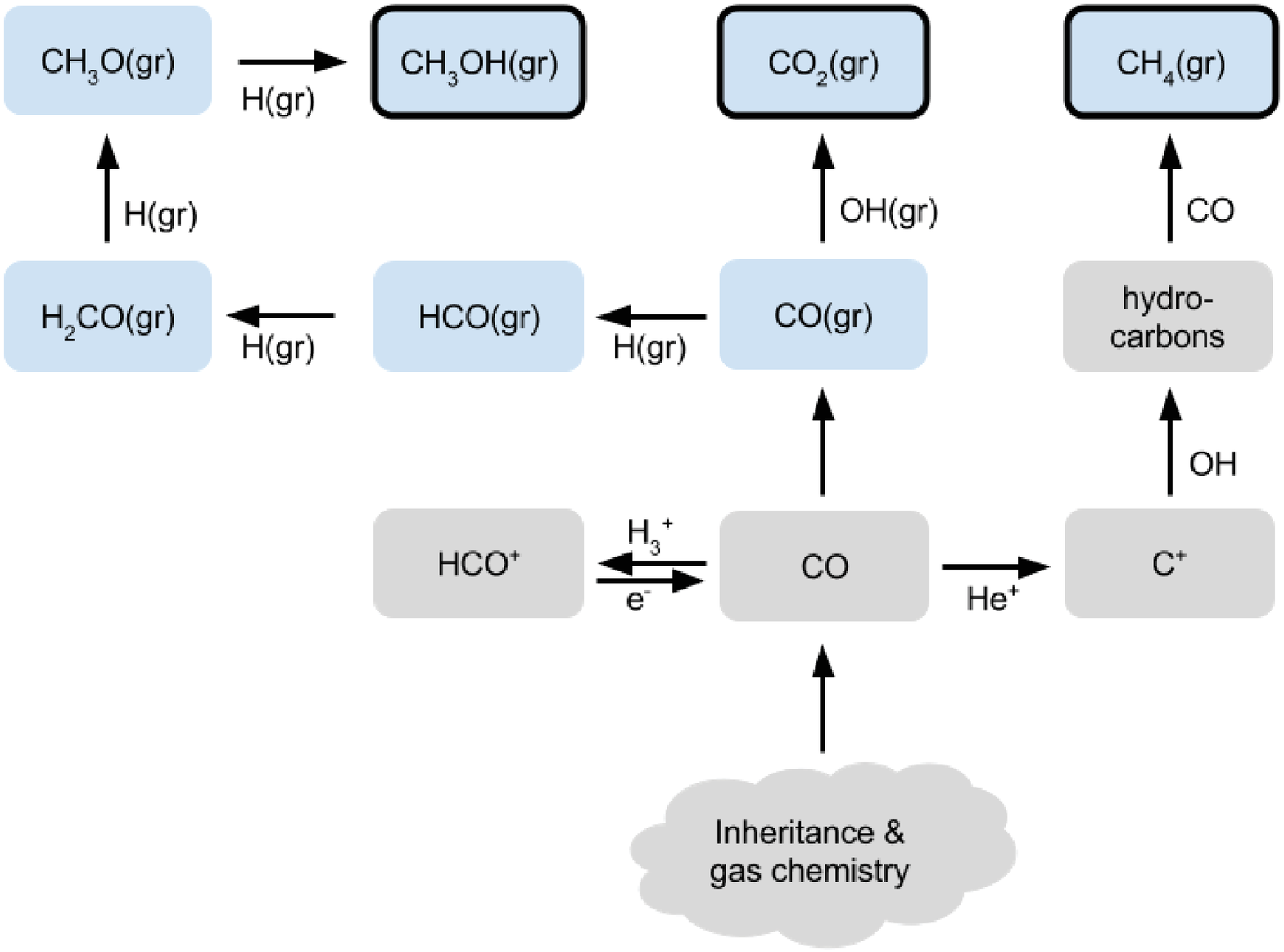}
\caption{Schematic of the chemical reactions relevant for CO reprocessing. Gas phase species are shown in grey. Ices are shown in blue. End state species are bolded. \label{schem}}
\end{figure*}

Figure~\ref{schem} provides a schematic of the most important chemical pathways via which CO is reprocessed.
In the warm molecular layer UV photons are able remove \water\ ice from the grain surface, creating gas phase OH and H. Both species are able to freeze out onto the grain surface. Once there they react with CO before thermal desorption can return CO to the gas phase. This slowly builds reservoirs of \cotwo\ ice and, to a lesser extent, \methanol\ ice, which are able to remain on the grain surface at temperatures where CO is primarily in the gas phase. 
This reprocessing does not occur for the warm models where the temperature has been increased by 20 K (warm and warm hiCR). In these warm models the temperature is high enough that even the less volatile species such as \cotwo\ cannot remain on the grain surface. 
Additionally, there is less CO reprocessing in models with more large grains. Large grains settle to the midplane, resulting in higher temperatures in the warm molecular layer as well as a higher flux of X-rays and UV photons (Figures~\ref{mtemp}-\ref{mxr}), all of which aid in keeping carbon in gas phase CO.

When a high cosmic ray rate is present, there is additional CO depletion on timescales of a few million years.
A high cosmic ray rate allows CO reprocessing to occur deeper in the disk due to the enhanced abundance of \hep\ and H$_3^+$. \hep\ dissociates CO, leading to a series of gas phase reactions which form hydrocarbons, primarily CH$_4$ (Figure~\ref{schem}). However, it is the reaction with \hhhp\ which ultimately plays a larger role in the reprocessing of CO. CO and \hhhp\ react to form HCO$^+$, which quickly reacts with an electron, returning the carbon to CO while also producing atomic hydrogen (Figure~\ref{schem}). Some of these H atoms freeze onto the grain surface, hydrogenating CO ice and eventually forming \methanol\ ice. Because of this process the warm hiCR models show roughly an order of magnitude of CO depletion by 3 Myr (Figure~\ref{summaryplotout}). \methanol, which has a higher binding energy than \cotwo, is able to remain frozen out even at the higher temperatures in the warm CR models. Because \methanol\ ice formation is less efficient than \cotwo\ ice formation, CO depletion takes longer in the warm hiCR models.

\subsection{Inner disk}
\begin{figure}[ht!]
\setlength{\intextsep}{0pt}
    \includegraphics[width=\columnwidth]{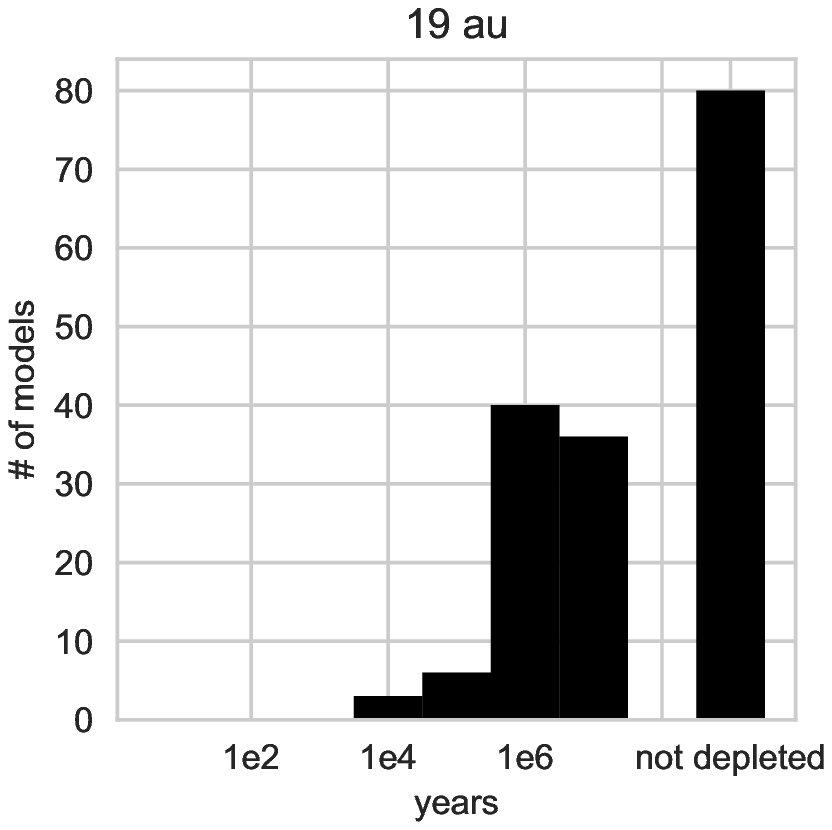}
\caption{Time at which the CO abundance first drops below \eten{-4} at 19 au for each model.\label{intimebar}}
\end{figure}

\begin{figure}[ht!]
\setlength{\intextsep}{0pt}
    \includegraphics[width=\columnwidth]{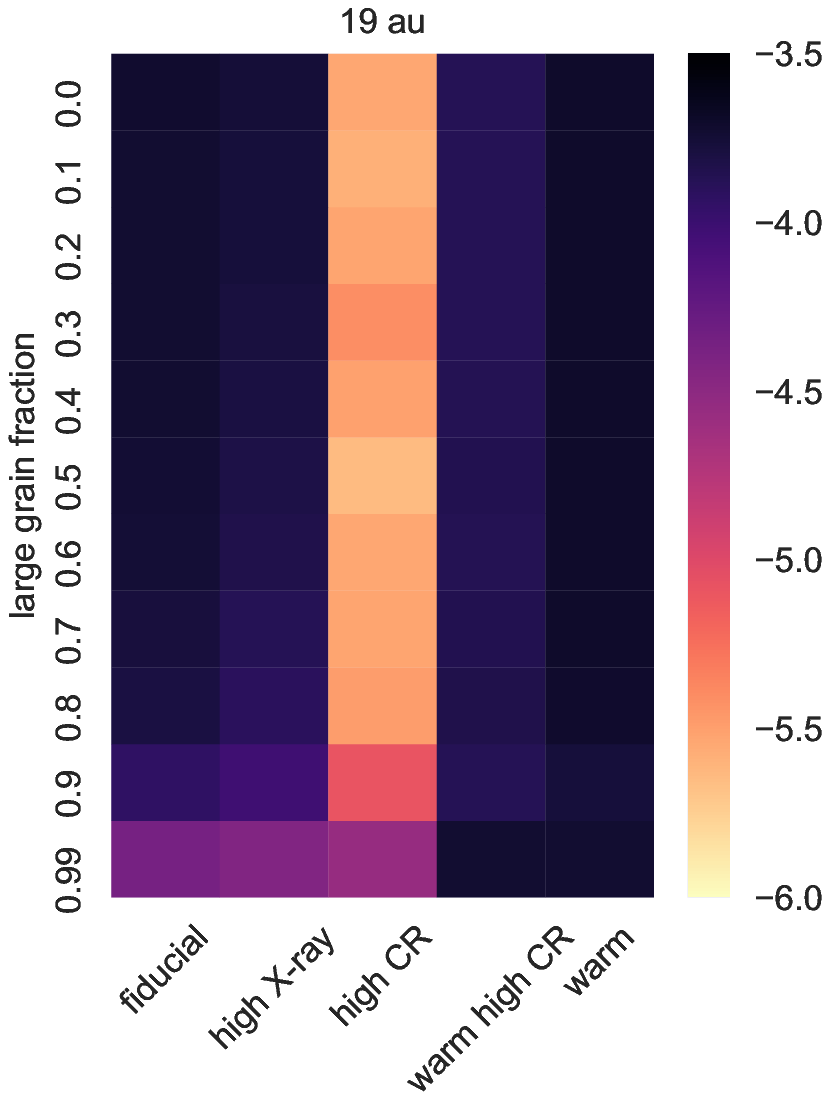}
\caption{Log CO abundance relative to \hh\ at 19 au for each model with a disk mass of 0.03 \msun. Additional figures for the 0.1 and 0.003 \msun\ disks can be found in Appendix \ref{appfigs}.\label{heatin}}
\end{figure}

At 19 au, interior to the midplane CO snow line, depletion timescales are longer with only 
5.5\% percent of models depleted after 0.1 Myr, 29.7\% depleted after 1 Myr, and 48.5\% depleted after 3 Myr (Figure~\ref{intimebar}).
The majority of models are not depleted in CO (Figure~\ref{heatin}). However, all of the hiCR models show CO depletion due to the increased abundance of \hhhp. Successive hydrogenation of CO ice reduces the CO column density by more than an order of magnitude within 1 Myr for the 0.1 and 0.03 \msun\ disks. 
As the fractional dust mass in large grains increases the disk becomes more settled, increasing the flux of UV photons in the warm molecular layer
This results in increased photodesorption of \water\ ice. As more oxygen becomes available to the chemistry H preferentially hydrogenates oxygen over CO. 
Ion induced CO depletion is also seen in the high X-ray luminosity models (hiXR) for the 0.003 \msun\ disk (Figure~\ref{summaryplotin}), in which the lower density allows the X-rays 
to penetrate deeper in the disk, producing \hhhp, though on longer timescales than those in the hiCR models (Figure~\ref{summaryplotin}).

The 0.1 and 0.03 \msun\ fiducial and hiXR models with 99\% of the dust mass in large grains are also able to chemically reprocess some CO due to the photodesorption of \water\ ice (Figures~\ref{heatin} \& \ref{largeheatin}). The resulting OH reacts with CO ice to form \cotwo\ ice.
At 19 au the 0.003 \msun\ disk is exposed to a substantially higher flux of UV photons than either the 0.1 or 0.03 \msun\ disk (Figure~\ref{smuv}). Thus, the production of \methanol\ in the hiCR models is hampered by the photodesorption of \water, regardless of the large grain fraction. When $\sim 60\%$ of the dust mass is in large grains the 0.003 \msun\ fiducial model shows some depletion of CO due to the photodesorption of \water\ ice (Figure~\ref{largeheatin}).
However, as the dust mass in large grains increases and the disk becomes more settled the increased photodesorption of \cotwo\ and \methanol\ makes it difficult for these species to remain on the grain and carbon is instead found primarily in gas phase CO regardless of the ionization structure (Figure~\ref{smallfiducialinner}). 

\subsection{Dominant carbon carriers}
For all models depleted in CO, carbon has become incorporated into less volatile species on the grain surface. Generally, \cotwo\ ice is the dominant carbon species, followed by \methanol\ ice. 
\cotwo\ can form directly from CO ice, while \methanol\ ice is formed via a series of hydrogenation reactions (Figure~\ref{schem}). Thus, the timescales needed to build a reservoir of \methanol\ ice are longer than those needed for \cotwo.
The atomic hydrogen needed to hydrogenate CO is an end product of \hh\ ionization.
As such, \methanol\ can only be the dominant carbon carrier in models with with a high cosmic ray rate (hiCR or warm hiCR).

It is important to note that both \cotwo\ ice and \methanol\ ice are end state species in our chemical network, meaning that the sequence of grain surface reactions which build progressively more complex species do not proceed past these two species. 
The absolute abundances of these ices in our models should not be taken as the expected \cotwo\ and \methanol\ ice abundances in observed systems. Rather, these ice abundances represent the total amount of carbon we expect to be locked up in ices. Reaction rates for grain surface chemistry involving more complex species are not currently well constrained, particularly for reactions involving \methanol\ \citep{Cuppen17}. Inclusion of these reactions in our network would be premature. 

\subsection{Reduced oxygen abundance}
\begin{figure}[h!]
\setlength{\intextsep}{0pt}
    \includegraphics[width=1.0\columnwidth]{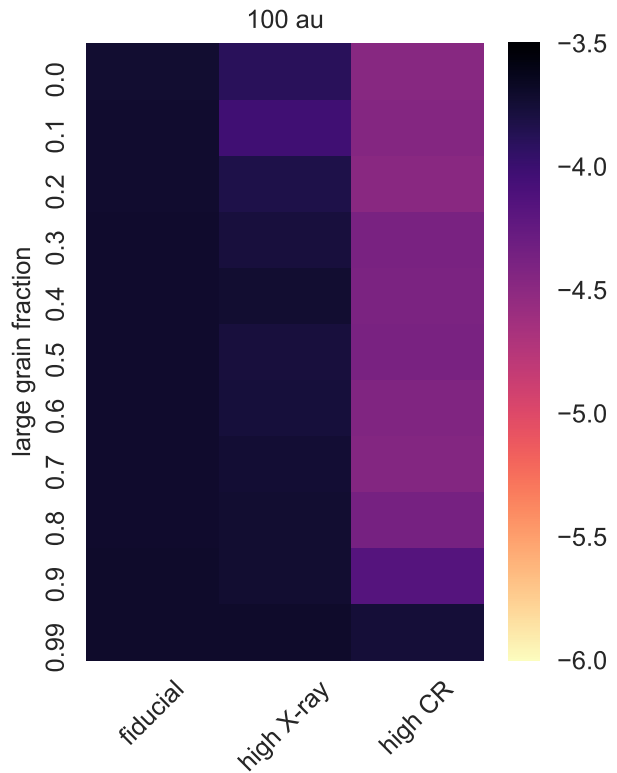}
  \caption{CO abundance relative to the initial abundance of \eten{-4} at 100 au for models with a reduced water/oxygen abundance. \label{outnowat}}
\end{figure}

\begin{figure}[h!]
\setlength{\intextsep}{0pt}
    \includegraphics[width=1.0\columnwidth]{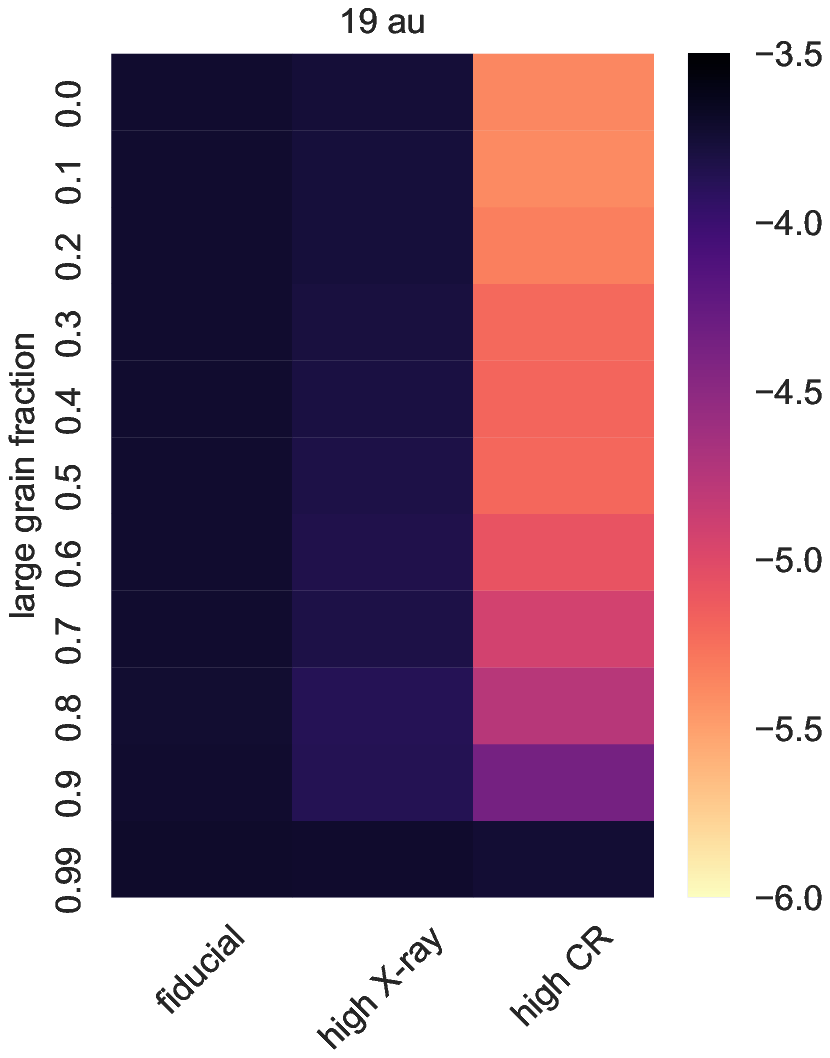}
  \caption{CO abundance relative to the initial abundance of \eten{-4} at 19 au for models with a reduced water/oxygen abundance. \label{nowaterin}}
\end{figure}

CO is able to be reprocessed into \cotwo\ because of the presence of water. However, observations suggest that the outer disks of many systems are depleted in water vapor \citep{Du15,Du17}. 
Additionally, \citet{Bergin16} find that the rings of hydrocarbon emission observed in TW Hya and DM Tau are best reproduced by models where C/O $>$ 1 in the gas.
To explore how a water and oxygen poor environment affects the volatile carbon chemistry we remove the initial water ice abundance and re-run the 0.03 \msun\ fiducial, hiCR, and hiXR models. The resulting CO abundances are shown in Figures \ref{outnowat} and \ref{nowaterin}. 
Of the 33 models water poor models, after 1 Myr  30.3\% are depleted in CO in the inner disk and 33.3\% are depleted in the outer disk. By 6 Myr this has risen to 57.6\% for the inner disk and 45.5\% for the outer disk. In comparison, for the corresponding models with a normal water abundance 42.4\% are depleted in the inner disk by 1 Myr, increasing to 69.7\% by 6 Myr, while in the outer disk all 33 models are depleted in CO by 1 Myr.
Without \water\ providing OH to convert CO into \cotwo\ it is much more difficult to remove carbon from CO. 
Unlike \cotwo\ formation, the \methanol\ formation pathway is not impeded by the lack of available oxygen. Some CO depletion is still able to take place, though on longer timescales than in the water rich models. 

In summary CO is chemically depleted in the outer disk warm molecular layer when the temperature is low enough for \cotwo\ to freeze out or there is a high cosmic ray rate. Additionally, more large grains lead to a more settled disk, resulting in higher temperatures and a higher UV flux in the warm molecular layer, which impede CO reprocessing. In the inner disk CO is chemically depleted when there is a high cosmic ray rate. Throughout the disk reducing the amount of oxygen available results in less reprocessing of CO.

\section{Discussion}\label{discussion}
Our models show that for an incident cosmic ray rate of $\sim \eten{-17}$ s$^{-1}$ CO is consistently destroyed and carbon is placed into ices, most commonly \cotwo\ and \methanol, and in some cases also CH$_{4}$, in agreement with previous studies \citep{Eistrup16,Yu16}. Reducing the ionization in the models impedes the reprocessing of CO. 
The modulation of cosmic rays by the solar wind as well as an inferred low cosmic ray ionization rate in TW Hya suggest that low cosmic ray rates due to modulation via winds should be common in disks \citep{Cleeves14}

However, in the outer disk warm molecular layer gas phase CO is reduced by an order of magnitude for the majority of our models, even those without a high cosmic ray rate. 
Furthermore, in our fiducial, low ionization models it is possible to reduce the CO abundance in the inner disk under certain circumstances.
In particular the disk must either have a low overall density, such as in our 0.003 \msun\ disk models, or the disk must be settled, with nearly all of the dust mass concentrated in the midplane. Under these circumstances X-ray ionization or UV photodesorption can drive CO reprocessing. The timescales for such reprocessing are typically longer than in the case of the cosmic ray driven chemistry, though order of magnitude depletion can occur on mega-year timescales given sufficient dust growth.

Recent high resolution continuum observations of young ($<1$ Myr) disks have revealed ringed dust features, which may be indicative of grain growth \citep{ALMA15,Zhang15,Pinte16}. 
Thus it is possible that disks are already highly settled by 1 Myr, allowing chemical depletion of CO to occur even without the presence of cosmic rays. 
Furthermore, there are many mechanisms, such as accretion onto the central star, photo-evaporation, and winds, which will, over time, reduce the density of the disk. Thus, disks for which chemical reprocessing is unable to occur at early times may still experience significant reprocessing at later times. 

There are a wide range of physical conditions which permit CO to be chemically reprocessed.
After 1 Myr, 99 out of 165 models have a CO abundance less than $\eten{-4}$ at 100 au.
This is sufficient to explain the CO abundances of order $\eten{-5}$ in many systems \citep{McClure16,Miotello17,Long17}.
However, the measured CO abundance in TW Hya is $\eten{-6}$, even interior to the midplane CO snowline \citep{Schwarz16,Zhang17}.
A high, ISM level, cosmic ray rate could result in the large depletion observed, at least in the inner disk, but is inconsistent with the ionization rate derived for this system \citep{Cleeves14}. 
Low gas-to-dust ratios based on CO observations in Lupus and Chamaeleon may indicate a two order of magnitude reduction of CO in other systems as well \citep{Miotello17,Long17}.
Additional mechanisms beyond chemical reprocessing are needed to explain these observed CO abundances.

CO to dust mass ratios below 100 may be the result of rapid gas loss, though the HD derived mass measurement for TW Hya suggest this is not the case in at least one system. 
Vertical mixing can preferentially place CO onto dust grains as ice, with the amount of freeze-out dependent on the level of turbulence \citep{Xu17}. Once on these grains it can be quickly reprocessed into less volatile species via grain surface reactions. Alternatively, if the small grains grow quickly, they will settle to the midplane, trapping some of the carbon bearing ices in the interior of large grains \citep{Krijt16b}.

It is important to note that the chemistry explored in this work is for a static disk model, with no mixing between vertical or radial zones. In reality, during the timescales over which CO is being reprocessed, grains are likely to grow, changing the physical conditions at a given location in the disk. Further, the inward drift of grains will transport ices to smaller radii, where they can sublimate. Radial dust drift will also expose the outer disk to higher temperatures and a higher photon flux, similar to the conditions in our warm models with 99\% of the dust mass in large grains. 
In the outer disk the typical timescale for CO reprocessing is less than 1 Myr.
If the timescales for dust growth and drift are longer than those for CO reprocessing there should still be substantial CO depletion in the outer disk, with carbon-bearing ices trapped in the interior of large grains and/or transported to the inner disk. If, however, reprocessing timescales are longer than those for dust growth and drift, the warmer temperatures will prevent substantial reprocessing of CO. Radial drift of large dust grains is expected to occur within a few hundred orbits \citep{Birnstiel16}. At 100 au around a 0.8 \msun\ star, such as that used in our models, one orbit takes approximately \eten{3} years which puts the drift timescale at several \eten{5} years, shorter than the typical CO reprocessing time in our models. However, for the majority of the fiducial, hiXR, and hiCR models half of the initial CO has been reprocessed by 0.1 Myr. Thus, chemical processing of CO in the outer disk should occur so long as the primary source of heating is the central star.

If 
disks observed to have a reduced CO gas abundance carry their carbon in \cotwo\ and \methanol\ ice, observations of these systems should show higher abundances of \cotwo\ and \methanol\ as the CO abundance decreases. In the inner few au of a disk, temperatures are warm enough for \cotwo\ to sublimate from grain surface ices. However, current observations of \cotwo\ emission at 15 \micron\ are best matched by models with \cotwo\ abundances in the range $\eten{-9}-\eten{-7}$ \citep{Bosman17}, well below the typical \cotwo\ ice abundance in our CO depleted models. This may imply that not all of the \cotwo\ ice has sublimated, perhaps because grain growth has trapped the ice in the interior of large planetesimals. Alternatively, there could be additional chemical processing of \cotwo\ ice beyond that currently considered by chemical networks. 
The James Webb Space Telescope will provide crucial insight into this problem by increasing the number of observations of \cotwo\ in disks. It will also provide additional information regarding the presence of methanol and other hydrocarbons, which may shed light on the issue of carbon evolution. 

\section{Summary}\label{summary}
We have run a grid of 198 chemical models, varying the disk mass, disk temperature, X-ray luminosity, cosmic ray ionization rate, grain size distribution, and initial water abundance 
in order to explore the conditions under which gas phase CO can be destroyed in protoplanetary disks.
\begin{enumerate}
\item At 100 au a majority of models (99 out of 165) with an ISM-like initial water abundances are depleted, with a CO abundance relative to \hh\ less than \eten{-4} after 1 Myr. Of these, only a minority of models (3) are extremely depleted with an abundance of less than \eten{-5}. The requirement for extreme depletion at this radius is the presence of cosmic rays. If cosmic rays are not present over a million years of evolution it is difficult to significantly deplete CO. However, in most cases after 6 Myr of evolution significant reductions in the CO abundance can occur. 
\item At 19 au 21 out of 165 models show extreme depletion in CO after 1 Myr. This extreme depletion only occurs in the presence of a high cosmic ray rate. After 6 Myr of evolution 45 models are extremely depleted while 77 models retain over half their initial CO.
\item Based on solar interaction with cosmic rays it is possible that cosmic rays are not abundant in young disk systems \citep[e.g.,][]{Cleeves13}. Thus, it is difficult for chemistry alone to produce the significant reductions in the warm molecular layer CO abundances needed to match current observational constrains. The one exception is low mass (0.003 \msun) disks, for which X-rays can provide the ionization needed to reprocess CO, though on timescales larger than the typical disk lifetime.

\item Removing water from the disk prior to CO depletion hampers the chemical reprocessing of CO.
\end{enumerate}

Therefore we conclude that chemistry alone is not responsible for the majority of CO depletion seen in disk systems. Other processes must be involved. Viscous mixing could be sending CO to deeper layers of the disk, where it is able to freeze onto grains which then grow. Additionally, carbon in the form of either CO of \cotwo\ ice could be locked inside large, many kilometer sized bodies, which are not easily evaporated within 1 Myr. 
In a subsequent work we will explore evolution of CO in the midplane, where conditions differ from those in the warm molecular layer. 
\\

This work was supported by funding from NSF grants
AST-1514670 and AST-1344133 (INSPIRE) as well as NASA NNX16AB48G. L.I.C. acknowledges the support of NASA
through Hubble Fellowship grant HST-HF2-51356.001.
K.Z. acknowledges the support of NASA through Hubble Fellowship grant HST-HF2-51401.001-A awarded by the Space Telescope Science Institute, which is operated by the Association of Universities for Research in Astronomy, Inc., for NASA, under contract NAS5-26555.


\begin{thebibliography}{48}
\expandafter\ifx\csname natexlab\endcsname\relax\def\natexlab#1{#1}\fi

\bibitem[{{Aikawa} \& {Herbst}(1999)}]{Aikawa99}
{Aikawa}, Y., \& {Herbst}, E. 1999, \aap, 351, 233

\bibitem[{{ALMA Partnership} {et~al.}(2015){ALMA Partnership}, {Brogan},
  {P{\'e}rez}, {Hunter}, {Dent}, {Hales}, {Hills}, {Corder}, {Fomalont},
  {Vlahakis}, {Asaki}, {Barkats}, {Hirota}, {Hodge}, {Impellizzeri}, {Kneissl},
  {Liuzzo}, {Lucas}, {Marcelino}, {Matsushita}, {Nakanishi}, {Phillips},
  {Richards}, {Toledo}, {Aladro}, {Broguiere}, {Cortes}, {Cortes}, {Espada},
  {Galarza}, {Garcia-Appadoo}, {Guzman-Ramirez}, {Humphreys}, {Jung}, {Kameno},
  {Laing}, {Leon}, {Marconi}, {Mignano}, {Nikolic}, {Nyman}, {Radiszcz},
  {Remijan}, {Rod{\'o}n}, {Sawada}, {Takahashi}, {Tilanus}, {Vila Vilaro},
  {Watson}, {Wiklind}, {Akiyama}, {Chapillon}, {de Gregorio-Monsalvo}, {Di
  Francesco}, {Gueth}, {Kawamura}, {Lee}, {Nguyen Luong}, {Mangum}, {Pietu},
  {Sanhueza}, {Saigo}, {Takakuwa}, {Ubach}, {van Kempen}, {Wootten},
  {Castro-Carrizo}, {Francke}, {Gallardo}, {Garcia}, {Gonzalez}, {Hill},
  {Kaminski}, {Kurono}, {Liu}, {Lopez}, {Morales}, {Plarre}, {Schieven},
  {Testi}, {Videla}, {Villard}, {Andreani}, {Hibbard}, \& {Tatematsu}}]{ALMA15}
{ALMA Partnership}, {Brogan}, C.~L., {P{\'e}rez}, L.~M., {et~al.} 2015, \apjl,
  808, L3

\bibitem[{{Ansdell} {et~al.}(2016){Ansdell}, {Williams}, {van der Marel},
  {Carpenter}, {Guidi}, {Hogerheijde}, {Mathews}, {Manara}, {Miotello},
  {Natta}, {Oliveira}, {Tazzari}, {Testi}, {van Dishoeck}, \& {van
  Terwisga}}]{Ansdell16}
{Ansdell}, M., {Williams}, J.~P., {van der Marel}, N., {et~al.} 2016, \apj,
  828, 46

\bibitem[{{Bai}(2016)}]{Bai16}
{Bai}, X.-N. 2016, \apj, 821, 80

\bibitem[{{Bergin} {et~al.}(2014){Bergin}, {Cleeves}, {Crockett}, \&
  {Blake}}]{Bergin14}
{Bergin}, E.~A., {Cleeves}, L.~I., {Crockett}, N., \& {Blake}, G.~A. 2014,
  Faraday Discussions, 168, 61

\bibitem[{{Bergin} {et~al.}(2016){Bergin}, {Du}, {Cleeves}, {Blake}, {Schwarz},
  {Visser}, \& {Zhang}}]{Bergin16}
{Bergin}, E.~A., {Du}, F., {Cleeves}, L.~I., {et~al.} 2016, \apj, 831, 101

\bibitem[{{Bergin} \& {Williams}(2017)}]{Bergin17}
{Bergin}, E.~A., \& {Williams}, J.~P. 2017, in Astrophysics and Space Science
  Library, Vol. 445, Astrophysics and Space Science Library, ed. M.~{Pessah} \&
  O.~{Gressel}, 1

\bibitem[{{Bergin} {et~al.}(2013){Bergin}, {Cleeves}, {Gorti}, {Zhang},
  {Blake}, {Green}, {Andrews}, {Evans}, {Henning}, {{\"O}berg}, {Pontoppidan},
  {Qi}, {Salyk}, \& {van Dishoeck}}]{Bergin13}
{Bergin}, E.~A., {Cleeves}, L.~I., {Gorti}, U., {et~al.} 2013, \nat, 493, 644

\bibitem[{{Bethell} \& {Bergin}(2011{\natexlab{a}})}]{Bethell11b}
{Bethell}, T.~J., \& {Bergin}, E.~A. 2011{\natexlab{a}}, \apj, 740, 7

\bibitem[{{Bethell} \& {Bergin}(2011{\natexlab{b}})}]{Bethell11a}
---. 2011{\natexlab{b}}, \apj, 739, 78

\bibitem[{{Birnstiel} {et~al.}(2016){Birnstiel}, {Fang}, \&
  {Johansen}}]{Birnstiel16}
{Birnstiel}, T., {Fang}, M., \& {Johansen}, A. 2016, \ssr

\bibitem[{{Bosman} {et~al.}(2017){Bosman}, {Bruderer}, \& {van
  Dishoeck}}]{Bosman17}
{Bosman}, A.~D., {Bruderer}, S., \& {van Dishoeck}, E.~F. 2017, \aap, 601, A36

\bibitem[{{Brickhouse} {et~al.}(2010){Brickhouse}, {Cranmer}, {Dupree}, {Luna},
  \& {Wolk}}]{Brickhouse10}
{Brickhouse}, N.~S., {Cranmer}, S.~R., {Dupree}, A.~K., {Luna}, G.~J.~M., \&
  {Wolk}, S. 2010, \apj, 710, 1835

\bibitem[{{Cleeves} {et~al.}(2013){Cleeves}, {Adams}, \& {Bergin}}]{Cleeves13}
{Cleeves}, L.~I., {Adams}, F.~C., \& {Bergin}, E.~A. 2013, \apj, 772, 5

\bibitem[{{Cleeves} {et~al.}(2014){Cleeves}, {Bergin}, {Alexander}, {Du},
  {Graninger}, {{\"O}berg}, \& {Harries}}]{Cleeves14}
{Cleeves}, L.~I., {Bergin}, E.~A., {Alexander}, C.~M.~O., {et~al.}
  2014, Science, 345, 1590

\bibitem[{{Cleeves} {et~al.}(2015){Cleeves}, {Bergin}, {Qi}, {Adams}, \&
  {{\"O}berg}}]{Cleeves15}
{Cleeves}, L.~I., {Bergin}, E.~A., {Qi}, C., {Adams}, F.~C., \& {{\"O}berg},
  K.~I. 2015, \apj, 799, 204

\bibitem[{{Cuppen} {et~al.}(2017){Cuppen}, {Walsh}, {Lamberts}, {Semenov},
  {Garrod}, {Penteado}, \& {Ioppolo}}]{Cuppen17}
{Cuppen}, H.~M., {Walsh}, C., {Lamberts}, T., {et~al.} 2017, \ssr

\bibitem[{{Draine} \& {Lee}(1984)}]{Draine84}
{Draine}, B.~T., \& {Lee}, H.~M. 1984, \apj, 285, 89

\bibitem[{{Du} {et~al.}(2015){Du}, {Bergin}, \& {Hogerheijde}}]{Du15}
{Du}, F., {Bergin}, E.~A., \& {Hogerheijde}, M.~R. 2015, \apjl, 807, L32

\bibitem[{{Du} {et~al.}(2017){Du}, {Bergin}, {Hogerheijde}, {van Dishoeck},
  {Blake}, {Bruderer}, {Cleeves}, {Dominik}, {Fedele}, {Lis}, {Melnick},
  {Neufeld}, {Pearson}, \& {Y{\i}ld{\i}z}}]{Du17}
{Du}, F., {Bergin}, E.~A., {Hogerheijde}, M., {et~al.} 2017, \apj, 842, 98

\bibitem[{{Dutrey} {et~al.}(1994){Dutrey}, {Guilloteau}, \& {Simon}}]{Dutrey94}
{Dutrey}, A., {Guilloteau}, S., \& {Simon}, M. 1994, \aap, 286

\bibitem[{{Eistrup} {et~al.}(2016){Eistrup}, {Walsh}, \& {van
  Dishoeck}}]{Eistrup16}
{Eistrup}, C., {Walsh}, C., \& {van Dishoeck}, E.~F. 2016, \aap, 595, A83

\bibitem[{{Favre} {et~al.}(2013){Favre}, {Cleeves}, {Bergin}, {Qi}, \&
  {Blake}}]{Favre13}
{Favre}, C., {Cleeves}, L.~I., {Bergin}, E.~A., {Qi}, C., \& {Blake}, G.~A.
  2013, \apjl, 776, L38

\bibitem[{{Fogel} {et~al.}(2011){Fogel}, {Bethell}, {Bergin}, {Calvet}, \&
  {Semenov}}]{Fogel11}
{Fogel}, J.~K.~J., {Bethell}, T.~J., {Bergin}, E.~A., {Calvet}, N., \&
  {Semenov}, D. 2011, \apj, 726, 29

\bibitem[{{Gleeson} \& {Axford}(1968)}]{Gleeson68}
{Gleeson}, L.~J., \& {Axford}, W.~I. 1968, \apj, 154, 1011

\bibitem[{{Harries}(2000)}]{Harries00}
{Harries}, T.~J. 2000, \mnras, 315, 722

\bibitem[{{Kama} {et~al.}(2016){Kama}, {Bruderer}, {Carney}, {Hogerheijde},
  {van Dishoeck}, {Fedele}, {Baryshev}, {Boland}, {G{\"u}sten}, {Aikutalp},
  {Choi}, {Endo}, {Frieswijk}, {Karska}, {Klaassen}, {Koumpia}, {Kristensen},
  {Leurini}, {Nagy}, {Perez Beaupuits}, {Risacher}, {van der Marel}, {van
  Kempen}, {van Weeren}, {Wyrowski}, \& {Y{\i}ld{\i}z}}]{Kama16}
{Kama}, M., {Bruderer}, S., {Carney}, M., {et~al.} 2016, \aap, 588, A108

\bibitem[{{Kastner} {et~al.}(1999){Kastner}, {Huenemoerder}, {Schulz}, \&
  {Weintraub}}]{Kastner99}
{Kastner}, J.~H., {Huenemoerder}, D.~P., {Schulz}, N.~S., \& {Weintraub}, D.~A.
  1999, \apj, 525, 837

\bibitem[{{Krijt} \& {Ciesla}(2016)}]{Krijt16}
{Krijt}, S., \& {Ciesla}, F.~J. 2016, \apj, 822, 111

\bibitem[{{Krijt} {et~al.}(2016){Krijt}, {Ciesla}, \& {Bergin}}]{Krijt16b}
{Krijt}, S., {Ciesla}, F.~J., \& {Bergin}, E.~A. 2016, \apj, 833, 285

\bibitem[{{Long} {et~al.}(2017){Long}, {Herczeg}, {Pascucci}, {Drabek-Maunder},
  {Mohanty}, {Testi}, {Apai}, {Hendler}, {Henning}, {Manara}, \&
  {Mulders}}]{Long17}
{Long}, F., {Herczeg}, G.~J., {Pascucci}, I., {et~al.} 2017, \apj, 844, 99

\bibitem[{{Mathis} {et~al.}(1977){Mathis}, {Rumpl}, \& {Nordsieck}}]{Mathis77}
{Mathis}, J.~S., {Rumpl}, W., \& {Nordsieck}, K.~H. 1977, \apj, 217, 425

\bibitem[{{McClure} {et~al.}(2016){McClure}, {Bergin}, {Cleeves}, {van
  Dishoeck}, {Blake}, {Evans}, {Green}, {Henning}, {{\"O}berg}, {Pontoppidan},
  \& {Salyk}}]{McClure16}
{McClure}, M.~K., {Bergin}, E.~A., {Cleeves}, L.~I., {et~al.} 2016, \apj, 831,
  167

\bibitem[{{McElroy} {et~al.}(2013){McElroy}, {Walsh}, {Markwick}, {Cordiner},
  {Smith}, \& {Millar}}]{McElroy13}
{McElroy}, D., {Walsh}, C., {Markwick}, A.~J., {et~al.} 2013, \aap, 550, A36

\bibitem[{{Miotello} {et~al.}(2014){Miotello}, {Bruderer}, \& {van
  Dishoeck}}]{Miotello14}
{Miotello}, A., {Bruderer}, S., \& {van Dishoeck}, E.~F. 2014, \aap, 572, A96

\bibitem[{{Miotello} {et~al.}(2017){Miotello}, {van Dishoeck}, {Williams},
  {Ansdell}, {Guidi}, {Hogerheijde}, {Manara}, {Tazzari}, {Testi}, {van der
  Marel}, \& {van Terwisga}}]{Miotello17}
{Miotello}, A., {van Dishoeck}, E.~F., {Williams}, J.~P., {et~al.} 2017, \aap,
  599, A113

\bibitem[{{Pinte} {et~al.}(2016){Pinte}, {Dent}, {M{\'e}nard}, {Hales}, {Hill},
  {Cortes}, \& {de Gregorio-Monsalvo}}]{Pinte16}
{Pinte}, C., {Dent}, W.~R.~F., {M{\'e}nard}, F., {et~al.} 2016, \apj, 816, 25

\bibitem[{{Raassen}(2009)}]{Raassen09}
{Raassen}, A.~J.~J. 2009, \aap, 505, 755

\bibitem[{{Reboussin} {et~al.}(2015){Reboussin}, {Wakelam}, {Guilloteau},
  {Hersant}, \& {Dutrey}}]{Reboussin15}
{Reboussin}, L., {Wakelam}, V., {Guilloteau}, S., {Hersant}, F., \& {Dutrey},
  A. 2015, \aap, 579, A82

\bibitem[{{Salyk} {et~al.}(2008){Salyk}, {Pontoppidan}, {Blake}, {Lahuis}, {van
  Dishoeck}, \& {Evans}}]{Salyk08}
{Salyk}, C., {Pontoppidan}, K.~M., {Blake}, G.~A., {et~al.} 2008, \apjl, 676,
  L49

\bibitem[{{Schwarz} {et~al.}(2016){Schwarz}, {Bergin}, {Cleeves}, {Blake},
  {Zhang}, {{\"O}berg}, {van Dishoeck}, \& {Qi}}]{Schwarz16}
{Schwarz}, K.~R., {Bergin}, E.~A., {Cleeves}, L.~I., {et~al.} 2016, \apj, 823,
  91

\bibitem[{{Smith} {et~al.}(2004){Smith}, {Herbst}, \& {Chang}}]{Smith04}
{Smith}, I.~W.~M., {Herbst}, E., \& {Chang}, Q. 2004, \mnras, 350, 323

\bibitem[{{Webber}(1998)}]{Webber98}
{Webber}, W.~R. 1998, \apj, 506, 329

\bibitem[{{Williams} \& {Best}(2014)}]{Williams14}
{Williams}, J.~P., \& {Best}, W.~M.~J. 2014, \apj, 788, 59

\bibitem[{{Xu} {et~al.}(2017){Xu}, {Bai}, \& {{\"O}berg}}]{Xu17}
{Xu}, R., {Bai}, X.-N., \& {{\"O}berg}, K. 2017, \apj, 835, 162

\bibitem[{{Youdin} \& {Lithwick}(2007)}]{Youdin07}
{Youdin}, A.~N., \& {Lithwick}, Y. 2007, \icarus, 192, 588

\bibitem[{{Yu} {et~al.}(2016){Yu}, {Willacy}, {Dodson-Robinson}, {Turner}, \&
  {Evans}}]{Yu16}
{Yu}, M., {Willacy}, K., {Dodson-Robinson}, S.~E., {Turner}, N.~J., \& {Evans},
  II, N.~J. 2016, \apj, 822, 53

\bibitem[{{Zhang} {et~al.}(2017){Zhang}, {Bergin}, {Blake}, {Cleeves}, \&
  {Schwarz}}]{Zhang17}
{Zhang}, K., {Bergin}, E.~A., {Blake}, G.~A., {Cleeves}, L.~I., \& {Schwarz},
  K.~R. 2017, Nature Astronomy, 1, 0130

\bibitem[{{Zhang} {et~al.}(2015){Zhang}, {Blake}, \& {Bergin}}]{Zhang15}
{Zhang}, K., {Blake}, G.~A., \& {Bergin}, E.~A. 2015, \apjl, 806, L7

\end{thebibliography}

\appendix
\section{Additional figures}\label{appfigs}

\begin{figure}[h!]
\setlength{\intextsep}{0pt}
    \includegraphics[width=0.5\columnwidth]{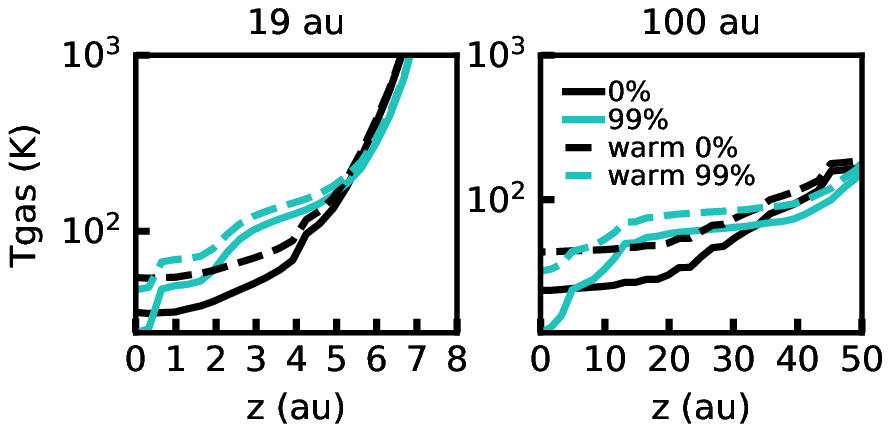}
\caption{Vertical gas temperature profiles at 19 au and 100 au for the 0.003 \msun\ disk. The 0\% large grain model is shown in black and the 99\% large grain model is shown in grey. Dashed lines indicate the temperature in the warm disk models. \label{smtemp}}
\end{figure}

\begin{figure}[h!]
\setlength{\intextsep}{0pt}
    \includegraphics[width=0.5\columnwidth]{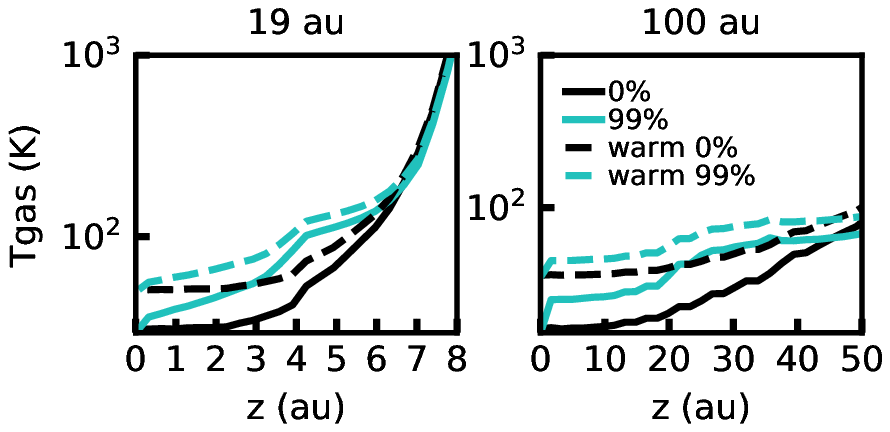}
\caption{Vertical gas temperature profiles at 19 au and 100 au for the 0.1 \msun\ disk. The 0\% large grain model is shown in black and the 99\% large grain model is shown in grey. Dashed lines indicate the temperature in the warm disk models. \label{lgtemp}}
\end{figure}

\begin{figure}[h!]
\setlength{\intextsep}{0pt}
    \includegraphics[width=0.5\columnwidth]{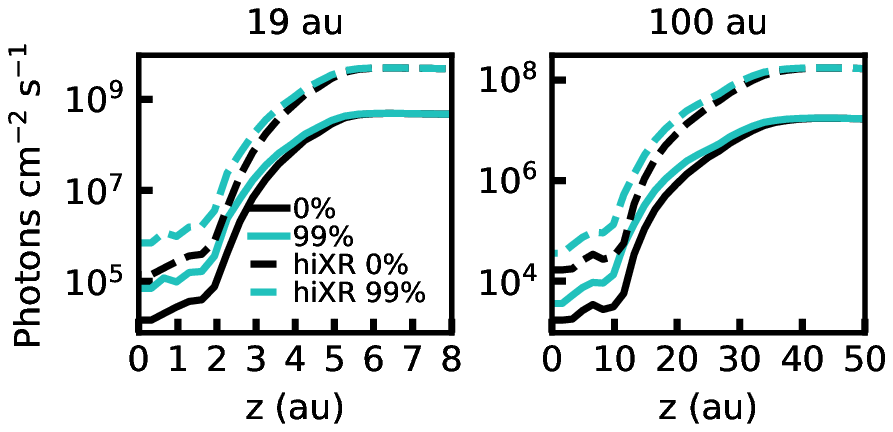}
\caption{Vertical profiles of the integrated X-ray flux at 19 au and 100 au for the 0.003 \msun\ disk. The 0\% large grain model is shown in black and the 99\% large grain model is shown in grey. Dashed lines indicate the temperature in the warm disk models. \label{smxr}}
\end{figure}

\begin{figure}[h!]
\setlength{\intextsep}{0pt}
    \includegraphics[width=0.5\columnwidth]{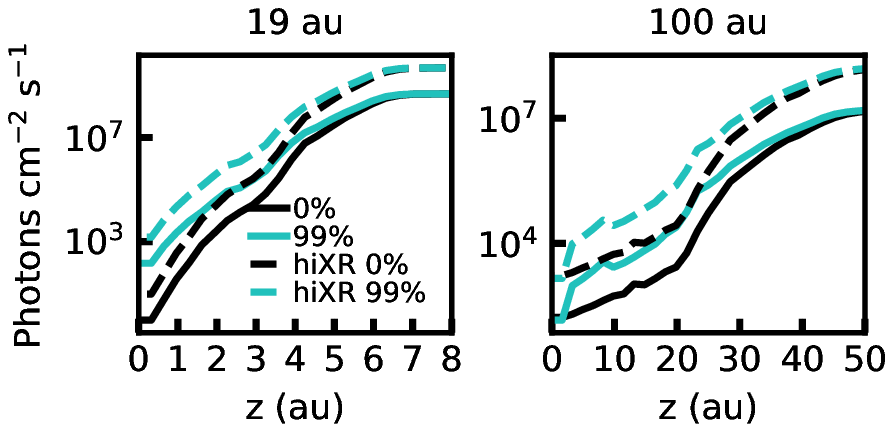}
\caption{Vertical profiles of the integrated X-ray flux at 19 au and 100 au for the 0.1 \msun\ disk. The 0\% large grain model is shown in black and the 99\% large grain model is shown in grey. Dashed lines indicate the temperature in the warm disk models. \label{lgxr}}
\end{figure}

\begin{figure}[h!]
\setlength{\intextsep}{0pt}
    \includegraphics[width=0.5\columnwidth]{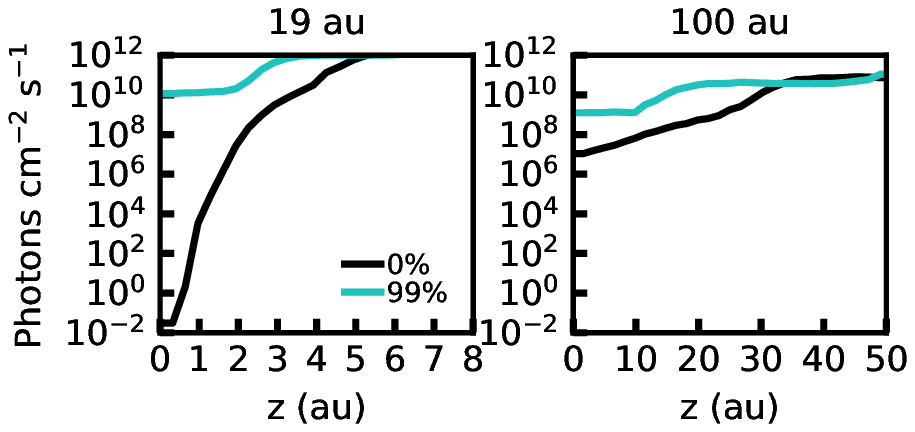}
\caption{Vertical profiles of the integrated UV flux at 19 au and 100 au for the 0.003 \msun\ disk. The 0\% large grain model is shown in black and the 99\% large grain model is shown in grey. Dashed lines indicate the temperature in the warm disk models. \label{smuv}}
\end{figure}

\begin{figure}[h!]
\setlength{\intextsep}{0pt}
    \includegraphics[width=0.5\columnwidth]{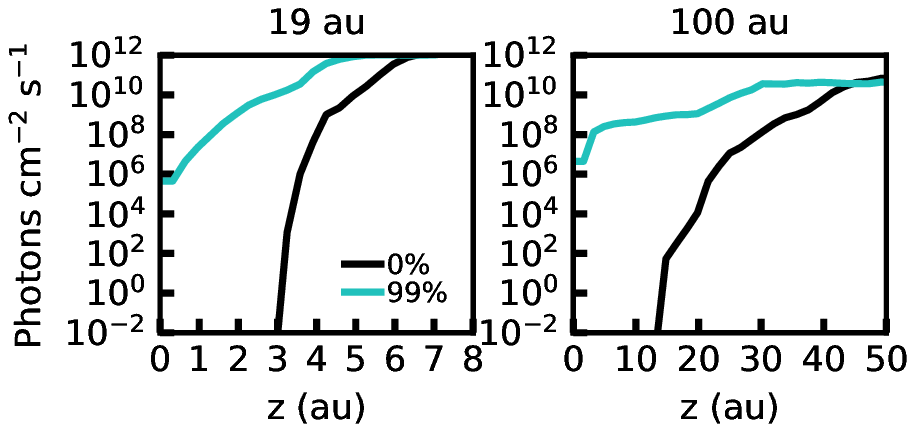}
\caption{Vertical profiles of the integrated UV flux at 19 au and 100 au for the 0.1 \msun\ disk. The 0\% large grain model is shown in black and the 99\% large grain model is shown in grey. Dashed lines indicate the temperature in the warm disk models. \label{lguv}}
\end{figure}

\begin{figure}[h!]
\setlength{\intextsep}{0pt}
    \includegraphics[width=0.5\columnwidth]{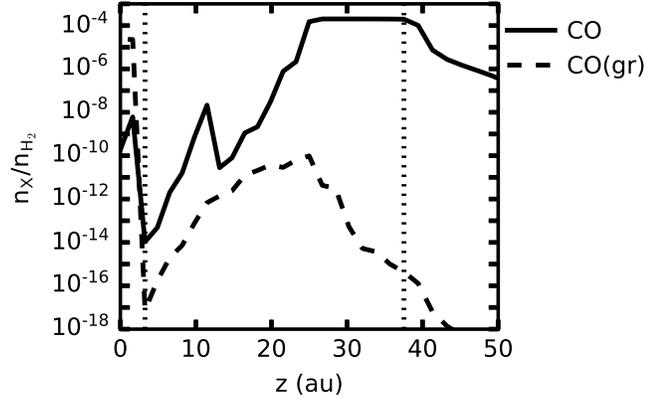}
\caption{CO gas and ice abundance as a function of height in the disk at 100 au for the fiducial 0.03 \msun\ model with 80\% large grains. The vertical dotted lines indicate the boundaries of the warm molecular layer. \label{wmldef}}
\end{figure}

\begin{figure}[ht!]
\setlength{\intextsep}{0pt}
    \includegraphics[width=1.0\textwidth]{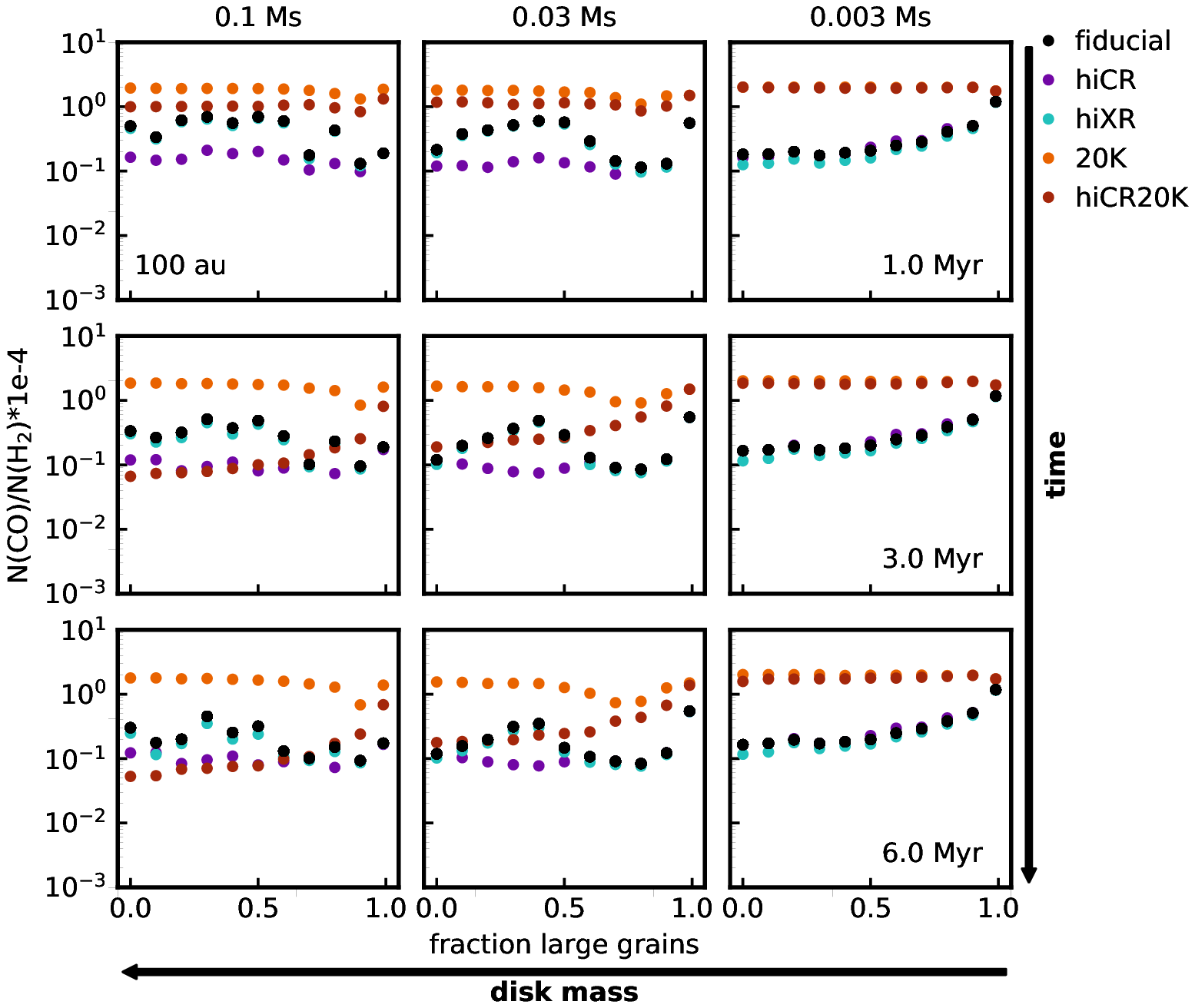}
  \caption{Predicted gas phase CO abundance in the warm molecular layer at 100 au for each model.\label{summaryplotout}}
\end{figure}

\begin{figure}[ht!]
\setlength{\intextsep}{0pt}
    \includegraphics[width=\columnwidth]{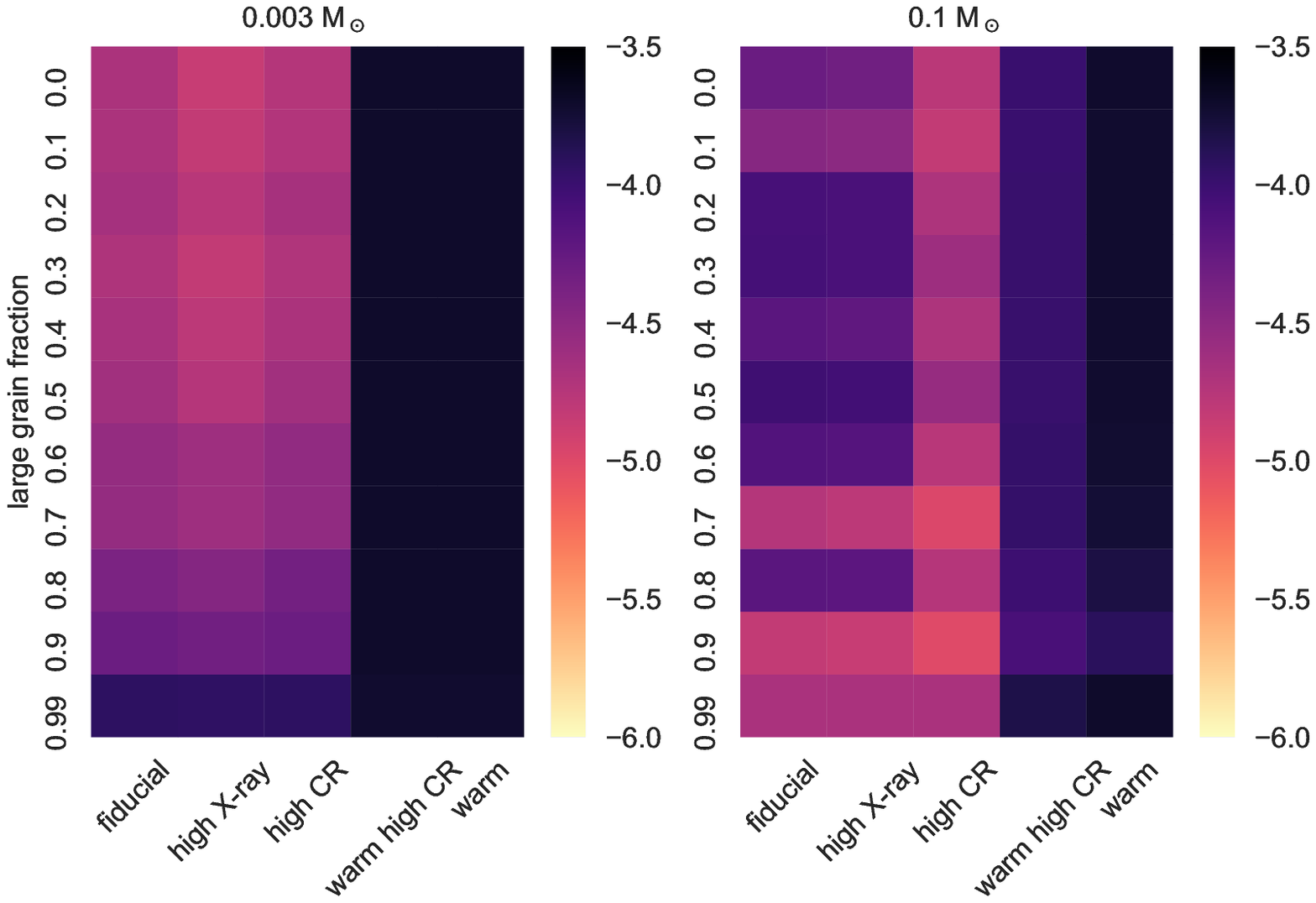}
\caption{Log CO abundance relative to \hh\ at 100 au for each model with disk masses of 0.003 and 0.1 \msun. \label{largeheatout}}
\end{figure}

\begin{figure}[ht!]
\setlength{\intextsep}{0pt}
    \includegraphics[width=0.5\columnwidth]{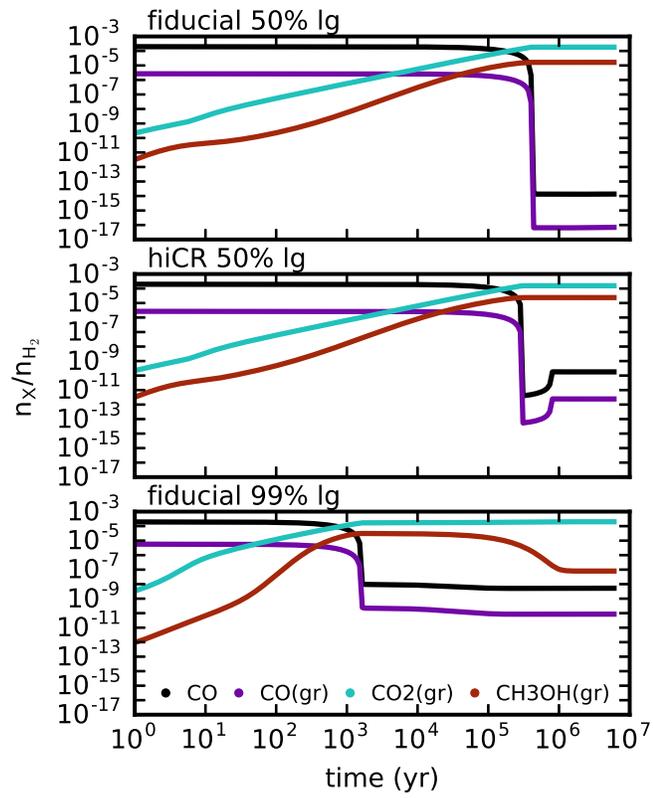}
\caption{Abundances of the most abundant carbon species as a function of time for select 0.03 \msun\ models at 100 au in radius and for a height of 11.5 au. \label{outwmlevo}}
\end{figure}

\begin{figure}
\setlength{\intextsep}{0pt}
    \includegraphics[width=1.0\textwidth]{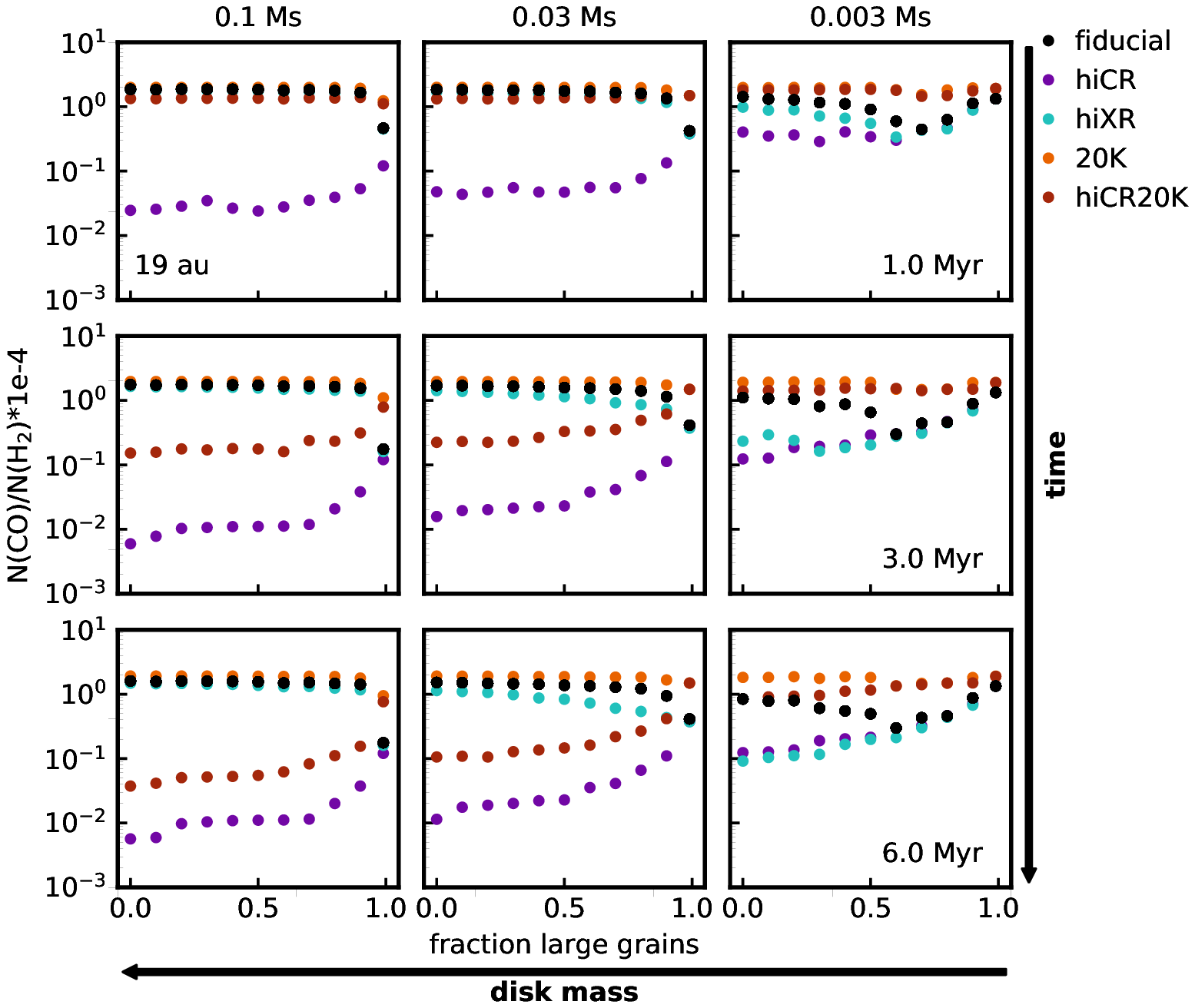}
  \caption{Predicted gas phase CO abundance in the warm molecular layer at 19 au for each model. \label{summaryplotin}}
\end{figure}

\begin{figure}[ht!]
\setlength{\intextsep}{0pt}
    \includegraphics[width=\columnwidth]{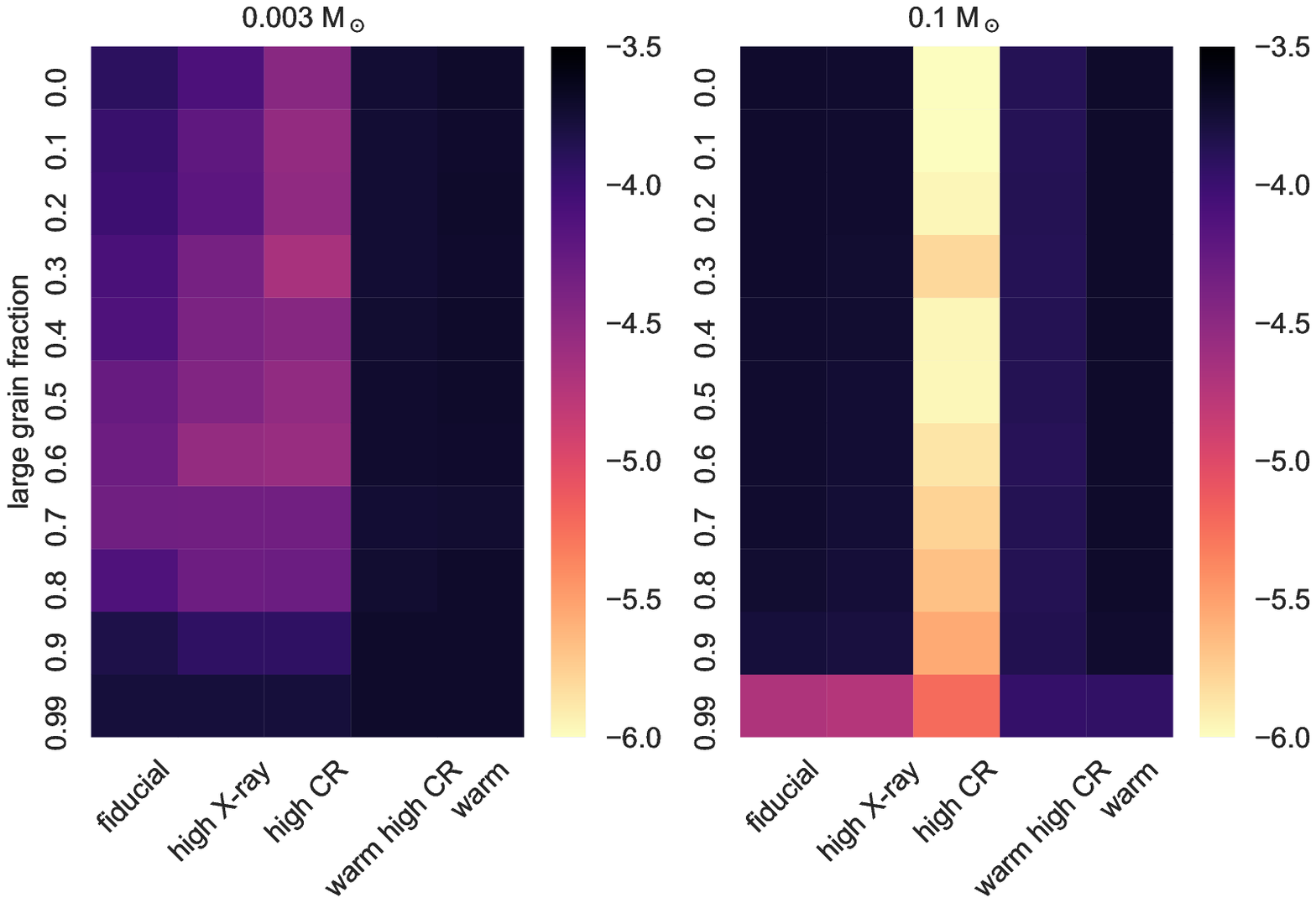}
\caption{Log CO abundance relative to \hh\  at 19 au for each model with disk masses of  0.003 and 0.1 \msun. \label{largeheatin}}
\end{figure}

\begin{figure}[ht!]
\setlength{\intextsep}{0pt}
    \includegraphics[width=0.5\columnwidth]{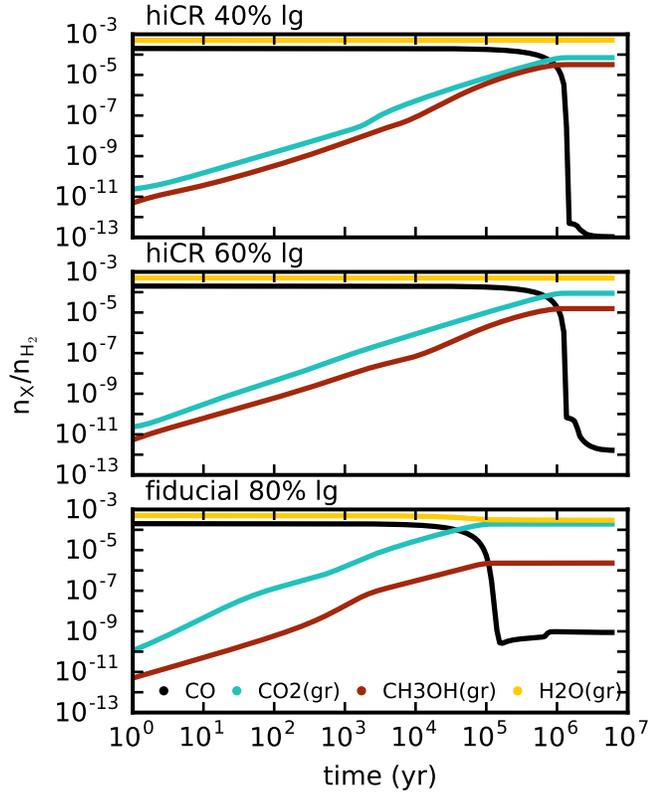}
\caption{Abundances as a function of time for select hiCR 0.03 \msun\ models at a radius of 19 au and a height of 2.3 au. \label{wmlgrevo}}
\end{figure}

\begin{figure}[ht!]
\setlength{\intextsep}{0pt}
    \includegraphics[width=0.5\columnwidth]{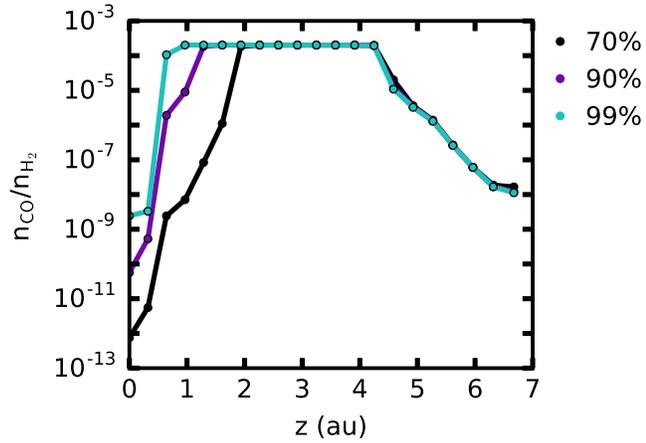}
\caption{CO abundance as a function of height in the disk at 19 au for the fiducial 0.003 \msun\ models with different large grain fractions. \label{smallfiducialinner}}
\end{figure}

\section{Abundance tables}\label{apptab}

\longrotatetable
\begin{deluxetable}{lrrrrrrrrrrrr}
\tabletypesize{\scriptsize}
\tablewidth{0pt}
\tablecaption{Top five most abundant carbon bearing species in the warm molecular layer at 100 au for each model after 1 Myr. Abundances are relative to \hh. Full table available online.}\label{tablewmlout}
\tablehead{
\colhead{Model} & \colhead{M$\mathrm{_{disk}}$} & \colhead{$f_l$ } & \colhead{Species} & \colhead{Abundance} & \colhead{Species} & \colhead{Abundance} & \colhead{Species} & \colhead{Abundance} & \colhead{Species} & \colhead{Abundance} & \colhead{Species} & \colhead{Abundance}  \\
}
\startdata
fiducial             & 0.003  & 0.0   & \cotwo(gr)    &    1.75e-04 & CO         &                      2.04e-05 & \methanol(gr)  &    4.37e-06 & CN(gr)     &    7.93e-07 & OCN(gr)                      &    4.18e-07 \\ 
fiducial             & 0.003  & 0.1   & \cotwo(gr)    &    1.73e-04 & CO         &                      2.06e-05 & \methanol(gr)  &    5.98e-06 & CN(gr)     &    7.32e-07 & OCN(gr)                      &    5.03e-07 \\ 
fiducial             & 0.003  & 0.2   & \cotwo(gr)    &    1.70e-04 & CO         &                      2.25e-05 & \methanol(gr)  &    7.35e-06 & CN(gr)     &    6.35e-07 & OCN(gr)                      &    5.74e-07 \\ 
fiducial             & 0.003  & 0.3   & \cotwo(gr)    &    1.73e-04 & CO         &                      1.97e-05 & \methanol(gr)  &    6.93e-06 & OCN(gr)    &    6.53e-07 & CN(gr)                       &    5.70e-07 \\ 
fiducial             & 0.003  & 0.4   & \cotwo(gr)    &    1.74e-04 & CO         &                      2.17e-05 & \methanol(gr)  &    4.41e-06 & OCN(gr)    &    7.03e-07 & CN(gr)                       &    4.60e-07 \\ 
fiducial             & 0.003  & 0.5   & \cotwo(gr)    &    1.75e-04 & CO         &                      2.33e-05 & \methanol(gr)  &    2.21e-06 & OCN(gr)    &    7.99e-07 & CH$_4$(gr)                      &    3.03e-07 \\ 
fiducial             & 0.003  & 0.6   & \cotwo(gr)    &    1.71e-04 & CO         &                      2.81e-05 & \methanol(gr)  &    8.04e-07 & OCN(gr)    &    7.73e-07 & CH$_4$(gr)                      &    3.43e-07 \\ 
fiducial             & 0.003  & 0.7   & \cotwo(gr)    &    1.72e-04 & CO         &                      2.82e-05 & OCN(gr)    &    8.36e-07 & \methanol(gr)  &    4.80e-07 & CH$_4$(gr)                      &    1.50e-07 \\ 
fiducial             & 0.003  & 0.8   & \cotwo(gr)    &    1.60e-04 & CO         &                      4.07e-05 & OCN(gr)    &    7.19e-07 & \methanol(gr)  &    1.41e-07 & C$^+$                           &    1.08e-07 \\ 
fiducial             & 0.003  & 0.9   & \cotwo(gr)    &    1.50e-04 & CO         &                      5.05e-05 & OCN(gr)    &    4.28e-07 & \methanol(gr)  &    2.22e-07 & C$^+$                           &    2.05e-07 \\ 
\enddata
\end{deluxetable}

\longrotatetable
\begin{deluxetable}{lrrrrrrrrrrrr}
\tabletypesize{\scriptsize}
\tablewidth{0pt}
\tablecaption{Top five most abundant carbon bearing species in the warm molecular layer at 19 au for each model after 1 Myr. Abundances are relative to \hh. Full table available online.}\label{tablewmlin}
\tablehead{
\colhead{Model} & \colhead{M$\mathrm{_{disk}}$} & \colhead{$f_l$ } & \colhead{Species} & \colhead{Abundance} & \colhead{Species} & \colhead{Abundance} & \colhead{Species} & \colhead{Abundance} & \colhead{Species} & \colhead{Abundance} & \colhead{Species} & \colhead{Abundance}  \\
}
\startdata
fiducial             & 0.003  & 0.0   & CO         &    1.21e-04 & \cotwo(gr)    &                      7.37e-05 & \methanol(gr)  &    8.96e-07 & HCN(gr)    &    7.52e-07 & HNC(gr)                      &    5.73e-07 \\ 
fiducial             & 0.003  & 0.1   & CO         &    1.03e-04 & \cotwo(gr)    &                      9.08e-05 & HCN(gr)    &    8.73e-07 & \methanol(gr)  &    8.42e-07 & HNC(gr)                      &    5.60e-07 \\ 
fiducial             & 0.003  & 0.2   & CO         &    9.81e-05 & \cotwo(gr)    &                      9.65e-05 & HCN(gr)    &    8.42e-07 & \methanol(gr)  &    6.94e-07 & HNC(gr)                      &    5.62e-07 \\ 
fiducial             & 0.003  & 0.3   & \cotwo(gr)    &    1.14e-04 & CO         &                      8.14e-05 & \methanol(gr)  &    8.35e-07 & HCN(gr)    &    7.54e-07 & HNC(gr)                      &    5.80e-07 \\ 
fiducial             & 0.003  & 0.4   & \cotwo(gr)    &    1.20e-04 & CO         &                      7.49e-05 & HCN(gr)    &    1.38e-06 & \methanol(gr)  &    8.77e-07 & HNC(gr)                      &    5.85e-07 \\ 
fiducial             & 0.003  & 0.5   & \cotwo(gr)    &    1.42e-04 & CO         &                      5.38e-05 & HCN(gr)    &    9.24e-07 & \methanol(gr)  &    6.30e-07 & HNC(gr)                      &    4.83e-07 \\ 
fiducial             & 0.003  & 0.6   & \cotwo(gr)    &    1.48e-04 & CO         &                      4.96e-05 & \methanol(gr)  &    1.88e-06 & HCN(gr)    &    7.54e-07 & HNC(gr)                      &    3.60e-07 \\ 
fiducial             & 0.003  & 0.7   & \cotwo(gr)    &    1.52e-04 & CO         &                      4.72e-05 & \methanol(gr)  &    1.15e-06 & HCN(gr)    &    6.60e-07 & HNC(gr)                      &    2.48e-07 \\ 
fiducial             & 0.003  & 0.8   & \cotwo(gr)    &    1.26e-04 & CO         &                      7.55e-05 & \methanol(gr)  &    1.68e-07 & HCN(gr)    &    1.33e-07 & OCN(gr)                      &    4.12e-08 \\ 
fiducial             & 0.003  & 0.9   & CO         &    1.48e-04 & \cotwo(gr)    &                      5.34e-05 & \methanol(gr)  &    9.93e-08 & OCN(gr)    &    1.93e-08 & HCN(gr)                      &    1.36e-08 \\ 
\enddata
\end{deluxetable}

\longrotatetable
\begin{deluxetable}{lrrrrrrrrrrrr}
\tabletypesize{\scriptsize}
\tablewidth{0pt}
\tablecaption{Top five most abundant carbon bearing species in the warm molecular layer at 19 au for each model after 1 Myr. Abundances are relative to \hh. Full table available online.}
\tablehead{
\colhead{Model} & \colhead{M$\mathrm{_{disk}}$} & \colhead{$f_l$ } & \colhead{Species} & \colhead{Abundance} & \colhead{Species} & \colhead{Abundance} & \colhead{Species} & \colhead{Abundance} & \colhead{Species} & \colhead{Abundance} & \colhead{Species} & \colhead{Abundance}  \\
}
\startdata
fiducial             & 0.003  & 0.0   & CO         &    1.37e-04 & \cotwo(gr)    &                      5.92e-05 & \methanol(gr)  &    7.14e-07 & HCN(gr)    &    5.47e-07 & HNC(gr)                      &    4.09e-07 \\ 
fiducial             & 0.003  & 0.1   & CO         &    1.27e-04 & \cotwo(gr)    &                      6.90e-05 & \methanol(gr)  &    6.86e-07 & HCN(gr)    &    6.29e-07 & HNC(gr)                      &    3.89e-07 \\ 
fiducial             & 0.003  & 0.2   & CO         &    1.24e-04 & \cotwo(gr)    &                      7.23e-05 & HCN(gr)    &    5.98e-07 & \methanol(gr)  &    5.82e-07 & HNC(gr)                      &    3.87e-07 \\ 
fiducial             & 0.003  & 0.3   & CO         &    1.21e-04 & \cotwo(gr)    &                      7.57e-05 & \methanol(gr)  &    7.18e-07 & HCN(gr)    &    5.11e-07 & HNC(gr)                      &    3.89e-07 \\ 
fiducial             & 0.003  & 0.4   & CO         &    1.18e-04 & \cotwo(gr)    &                      7.85e-05 & HCN(gr)    &    8.83e-07 & \methanol(gr)  &    6.94e-07 & HNC(gr)                      &    3.95e-07 \\ 
fiducial             & 0.003  & 0.5   & CO         &    1.04e-04 & \cotwo(gr)    &                      9.33e-05 & \methanol(gr)  &    6.34e-07 & HCN(gr)    &    5.97e-07 & HNC(gr)                      &    3.24e-07 \\ 
fiducial             & 0.003  & 0.6   & \cotwo(gr)    &    1.01e-04 & CO         &                      9.53e-05 & \methanol(gr)  &    2.98e-06 & HCN(gr)    &    4.94e-07 & HNC(gr)                      &    2.51e-07 \\ 
fiducial             & 0.003  & 0.7   & \cotwo(gr)    &    1.10e-04 & CO         &                      8.87e-05 & \methanol(gr)  &    1.83e-06 & HCN(gr)    &    4.55e-07 & HNC(gr)                      &    1.86e-07 \\ 
fiducial             & 0.003  & 0.8   & CO         &    1.07e-04 & \cotwo(gr)    &                      9.36e-05 & \methanol(gr)  &    2.40e-07 & HCN(gr)    &    1.32e-07 & C$^+$                           &    1.12e-07 \\ 
fiducial             & 0.003  & 0.9   & CO         &    1.50e-04 & \cotwo(gr)    &                      5.17e-05 & \methanol(gr)  &    1.64e-07 & C$^+$         &    1.51e-07 & OCN(gr)                      &    3.28e-08 \\ 
\enddata
\end{deluxetable}

\end{document}